\documentclass[aps,prd,twocolumn,showpacs,floatfix,preprintnumbers,amsmath,amssymb,nofootinbib,groupedaddress, groupedaddress, notitlepage]{revtex4-1}

\input epsf
\usepackage{graphicx}
\usepackage{color}
\usepackage{mathtools}
\usepackage{mathbbol}

\newcommand{\beq}{\begin{equation}}
\newcommand{\eeq}{\end{equation}}
\newcommand{\barr}{\begin{eqnarray}}
\newcommand{\earr}{\end{eqnarray}}

\newcommand{\rme}{\textrm{e}}

\newcommand{\bs}{\boldsymbol}

\newcommand{\apjl}{Astrophys. J. Lett.}
\newcommand{\aap}{Astron. Astrophys.}
\newcommand{\apjs}{Astrophys. J. Suppl. Ser.}

\newcommand{\mnras}{Mon. Not. R. Astron. Soc.}

\newcommand{\physrep}{Phys. Rep.}

\newcommand{\jcap}{J. Cosm. Astropart. Phys.}

\usepackage{color}

\newcommand{\change}[1]{{#1}}
\newcommand{\changee}[1]{{#1}}
\newcommand{\revision}[1]{{#1}}

\newcommand{\lsim}{\mathrel{\hbox{\rlap{\lower.55ex\hbox{$\sim$}} \kern-.3em \raise.4ex \hbox{$<$}}}}
\newcommand{\gsim}{\mathrel{\hbox{\rlap{\lower.55ex\hbox{$\sim$}} \kern-.3em \raise.4ex \hbox{$>$}}}}

\begin{document}
\title{Cosmic microwave background limits on accreting primordial black holes}
\author{Yacine Ali-Ha\"imoud and Marc Kamionkowski}
\affiliation{Department of Physics and Astronomy, Johns Hopkins University, Baltimore, MD 21218, USA}
\date{\today}

\begin{abstract}
Interest in the idea that primordial black holes (PBHs) might
comprise some or all of the dark matter has recently been rekindled
following LIGO's first direct detection of a binary-black-hole
merger. Here we revisit the effect of accreting PBHs on the cosmic
microwave background (CMB) frequency spectrum and
angular temperature/polarization power spectra. We compute the accretion
rate and luminosity of PBHs, accounting for their suppression by
Compton drag and Compton cooling by CMB photons. We estimate the gas
temperature near the Schwarzschild radius, and hence the free-free
luminosity, accounting for the cooling resulting from
collisional ionization when the background gas is mostly
neutral. We account approximately for the velocities of PBHs
with respect to the background gas. \change{We provide a simple analytic estimate of the efficiency of energy deposition in the plasma.} We find that the spectral
distortions generated by accreting PBHs are too small to be
detected by FIRAS\change{, as well as by} future experiments now being considered. \change{We analyze \emph{Planck} CMB temperature and polarization data and find, under our most conservative hypotheses, \changee{and at the order-of-magnitude level}, that they rule out PBHs with masses $\gtrsim 10^2 \,M_{\odot}$ as the dominant component of dark matter.}
%
\end{abstract}

\maketitle

\section{Introduction}

\change{The idea of primordial black holes (PBHs) was first put forward by Zel'dovich and Novikov in the
sixties \cite{Zeldovich_67}. Developing it further, Hawking argued that
early-Universe fluctuations could lead to the formation of PBHs
with masses down to the Planck mass \cite{Hawking_71}. Chapline
was the first to suggest that PBHs could make the dark matter
(DM) \cite{Chapline_75}. Though this class of DM candidate has taken a
back seat to the notion that DM is a new elementary
particle \cite{Jungman_96,Bertone:2004pz,Bergstrom:2000pn,Hooper:2007qk,Zurek:2013wia}, the idea of PBH dark matter was recently rekindled \cite{Bird_16, Sasaki_16}, following the first detection of two merging $\sim30\, M_\odot$ black holes by LIGO \cite{LIGO_16}. 
Given the increasingly constraining null searches for particle DM, PBHs and their observational consequences are worth reconsidering \cite{Carr_10,Carr_16}.}

\change{The abundance of PBHs is constrained by a variety of observations in several mass ranges (for a comprehensive review see Refs.~\cite{Carr_10, Carr_16}). To cite only a few constraints, null microlensing searches exclude compact objects with masses $\lesssim10 \, M_\odot$ \cite{Macho_01, Eros_07}, and wide-binary surveys exclude those with masses $\gtrsim10^2 \,M_\odot$ \cite{Quinn_09, Monroy_14}. For PBHs more massive than $\sim 1\, M_{\odot}$, strong constraints were derived by Ricotti, Ostriker, and Mack \cite{Ricotti_08} (hereafter ROM) from the cosmic microwave background (CMB) frequency spectrum
and temperature and polarization anisotropies.} The basic idea behind these
limits is that PBHs accrete primordial gas in the early Universe
and then convert a fraction of the accreted mass to
radiation. The \change{resulting injection of energy into the primordial plasma then affects its} thermal and ionization
histories \cite{Carr_81}, and thus \change{leads to distortions} to the frequency spectrum of the CMB
and to its temperature/polarization power spectra.
ROM estimate that CMB anisotropy measurements by WMAP
\cite{Spergel_07} and limits on CMB spectral distortions by
FIRAS \cite{Fixen_96} exclude PBHs with masses $M \gtrsim
1\, M_{\odot}$ and $M \gtrsim 0.1 \, M_{\odot}$, respectively, as
the dominant component of dark matter. \change{Using ROM's results, Ref.~\cite{Chen:2016pud} strengthened these constraints with \emph{Planck} data.} Here we re-examine in detail CMB limits to the PBH
abundance, building on and expanding the work of ROM.

It is notoriously difficult to estimate from first principles
and self-consistently the accretion rate onto a central object
and the corresponding radiative efficiency (see, e.g., the
discussion in Chapter 14 of Ref.~\cite{ST_83}). In this work, we
strive to estimate the minimum physically-plausible PBH
luminosity in order to set the most conservative constraints to
the PBH abundance. 
\change{The bounds we derive are significantly weaker than
those of ROM: using Planck temperature and polarization data \cite{Planck_16}, we find that only PBHs with masses $M \gtrsim 10^2\, M_{\odot}$ can be conservatively excluded as the dominant component of the dark matter. Moroever, we find that CMB spectral-distortion measurements, both current and upcoming, do not place any constraints on PBHs. }

The single largest difference between our work and ROM's lies in
the adopted radiative efficiency $\epsilon \equiv L/\dot{M}
c^2$ to convert the mass accretion rate $\dot{M}$ to luminosity $L$. For masses $M \lesssim 10^4 \, M_{\odot}$, both ROM
and this work conservatively assume a quasi-spherical accretion
flow.  Shapiro \cite{Shapiro_73} provided a first-principles
estimate of the radiative efficiency for this problem.  This
shows that $\epsilon \propto
\dot{m} \equiv \dot{M} c^2/L_{\rm Edd}$, where $L_{\rm Edd}$ is
the Eddington luminosity. While ROM assume a fixed
$\epsilon/\dot{m} =0.011$ for $\dot{m} \leq 1$, we
generalize Shapiro's calculation, in particular accounting for
Compton cooling by ambient CMB photons, and explicitly compute
$\epsilon/\dot{m}$ as a function of PBH mass and redshift. We
find that $\epsilon/\dot{m}$ never exceeds $\sim 10^{-3}$
(corresponding to Shapiro's result for accretion from an HII
region), and can be as low as $\sim 10^{-5}$ after recombination
(corresponding to Shapiro's result for accretion from an HI
region), or even lower at high redshifts and for large PBH
masses for which Compton cooling is important. A few other
differences moreover contribute to lowering the mass accretion
rate with respect to that derived by ROM, as detailed in the
remainder of this article.

The rest of this paper is organized as follows. The core of our
calculation is laid out in Section \ref{sec:accretion}: there we
compute the accretion rate and luminosity of an accreting black
hole in the early Universe. \revision{We discuss the local feedback of this radiation in Section \ref{sec:feedback}}. In Section \ref{sec:deposition} we
estimate the efficiency with which the energy injected by PBHs
is deposited into the plasma. We then estimate the effect of
PBHs on CMB observables and derive the resulting constraints in
Section \ref{sec:cmb}. Finally, we conclude in Section
\ref{sec:conclusion}. To keep the calculation tractable
analytically we must make several approximations and
assumptions. In order to not disrupt the flow of the
calculation, we defer the verification of these assumptions to
Appendix \ref{app: consistency}. In Appendix
\ref{app:comparison} we compare our analytic approximation for
the efficiency of energy deposition in the plasma to existing
studies.

\section{Accretion onto a black hole in the early Universe} \label{sec:accretion}

\subsection{General considerations and calculation outline}

The first aspect to consider is the geometry of the
accretion. If the characteristic angular momentum of the
accreted gas (at the Bondi radius) is smaller than the angular
momentum at the innermost stable circular orbit, the accretion
is mostly spherical. Otherwise, an accretion disk forms. Disk
accretion is typically much more efficient than spherical
accretion at converting accreted mass into radiation. Indeed,
while in the latter case the dominant source of luminosity is
bremsstrahlung radiation from the hot ionized plasma near the
event horizon, in the former case the large viscous heating
required to dissipate angular momentum leads to radiating a
significant fraction of the rest-mass energy \cite{ST_83}. It is
difficult to estimate the angular momentum of the gas accreting
onto PBHs, as it requires knowledge of the PBH-baryon relative
velocity on scales of the order of the Bondi radius, which is
much smaller than any currently observed cosmological scale. A
correct estimate of this relative velocity would moreover
require accounting for the (non-linear) clustering of
PBHs. Following our philosophy to derive the most conservative
and physically-motivated accretion rate and luminosity, we shall
therefore adopt a spherical accretion model, expanding on the
classic work of Shapiro \cite{Shapiro_73}. We note that this is
also the underlying assumption made in ROM for PBH masses $M
\lesssim 10^3 -10^4\, M_{\odot}$ (see their Section 3.3). 

Another difficulty is that of local feedback. The radiation
emanating from the accreting PBH may indeed ionize and/or heat
the accreting gas, which would in turn affect the radiative
output. \revision{We show} in Appendix \ref{app:feedback} that
thermal feedback is negligible for all masses and redshifts
considered (consistent with ROM's Section 4.2.1 results). \revision{We will also see in Section \ref{sec:ion_feedback} that the Str\"omgren radius is always significantly smaller than the Bondi radius (consistent with our luminosity being significantly lower than that of ROM, who find that the photoionized region is marginally smaller than the Bondi radius). Hence one can assume that in the outermost region of the accretion flow, the ionization fraction is approximately equal to the background value. Close enough to the black hole, the gas eventually becomes fully ionized, either through photoionizations by the outgoing radiation field, or collisional ionizations, or both. We will see in Section \ref{sec:ion_feedback} that neither ionization process clearly prevails. To circumvent a complex self-consistent calculation of the luminosity and ionization profile, we shall consider the two limiting cases where one of the two ionization processes is dominant, and quote our results for both. In the first case, we shall completely neglect any radiative
feedback, and assume that the ionization fraction $x_e$ is equal
to the background value $\overline{x}_e$, until the temperature
of the gas reaches $ \sim 10^4$ K, at which point the gas gets
collisionally ionized. In our second limiting case, we assume
that the radiation from the PBH photoionizes the gas up to a radius beyond that at which $T \sim 10^4$ K (so collisional ionizations are not relevant), yet inside the Bondi radius. In all figures we refer to the former case by \emph{collisional ionization} and the latter by \emph{photoionization}. The correct result (within our overall model) lies somewhere between these two limiting cases. The difference between the final results in the two limits illustrates the relatively large theoretical uncertainty associated with this calculation.}

\begin{figure}[h]
\includegraphics[width = \columnwidth]{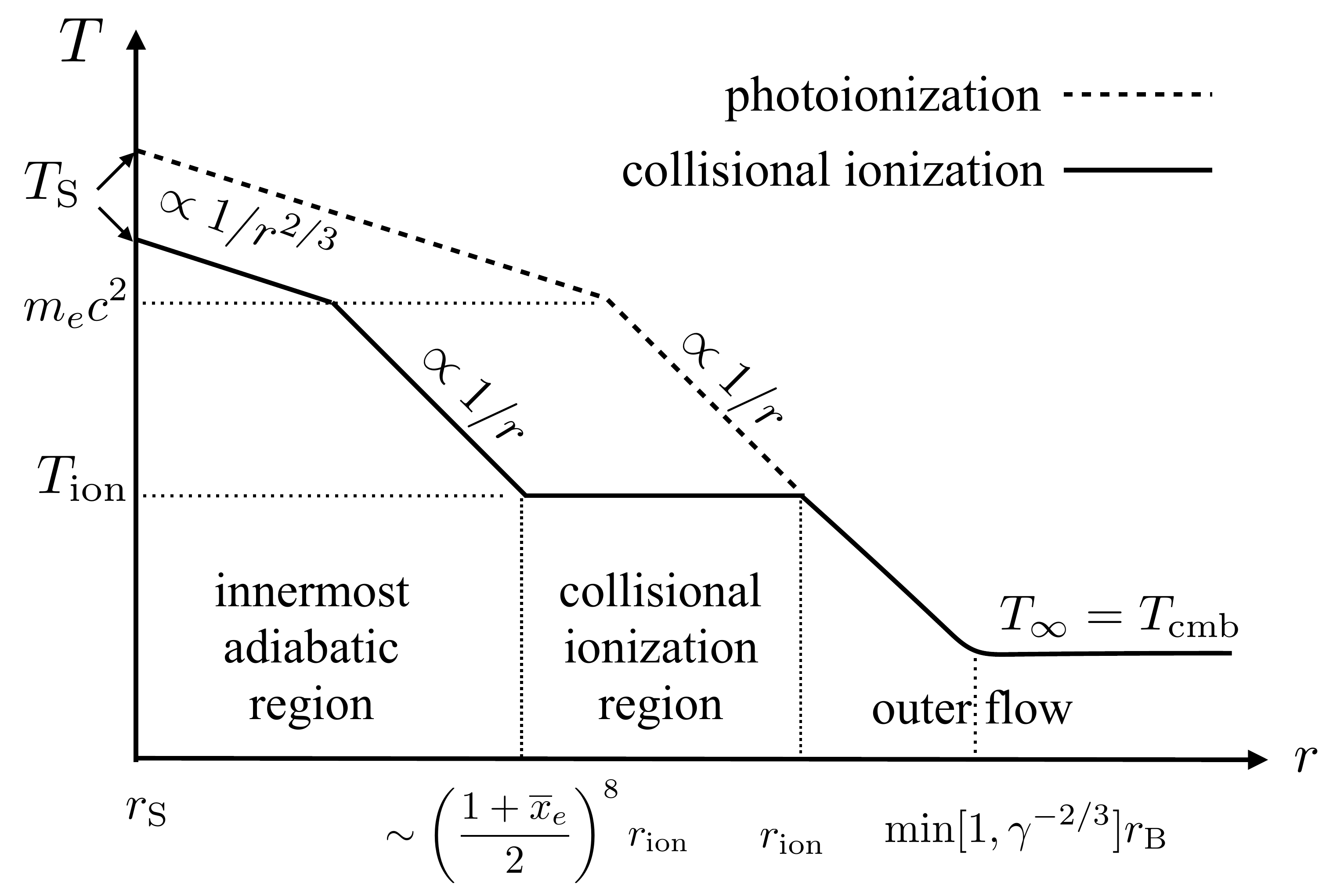}
\caption{Schematic temperature profile for the gas accreting onto a BH. If Compton cooling is efficient ($\gamma \gg 1$), the gas temperature remains close to the CMB temperature down to $r \sim \gamma^{-2/3} r_{\rm B}$, where $r_{\rm B}$ is the Bondi radius. The temperature then increases adiabatically as $T \propto \rho^{2/3} \propto 1/r$. \revision{If photoionizations can be neglected}, and if the background gas is partially neutral, the gas gets collisionally ionized at nearly constant temperature once it reaches $T_{\rm ion} \approx 1.5 \times 10^4$ K. Once the gas is fully ionized, the temperature resumes increasing adiabatically as $T \propto 1/r$ until electrons become relativistic, at which point the change in the adiabatic index implies $T \propto \rho^{4/9} \propto r^{-2/3}$. \revision{If the luminosity of the accreting PBH is large enough, the gas is photoionized instead of collisionally ionized. In that case the gas temperature reaches larger values near the black hole horizon.}} \label{fig:T_of_r}
\end{figure}

In what follows we split the calculation of the hydrodynamical
and thermal state of the gas accreting onto a BH into three
regions. First, in Section \ref{sec:outer}, we study the
outermost region where we assume a constant free-electron
fraction \revision{$x_e$ equal to the background value $\overline{x}_e$}. We solve for the
steady-state fluid and heat equations, as well as the accretion
rate, accounting for Compton drag (as in
Ref.~\cite{Ricotti_07}).  We also include, for the first time in
this context, for Compton cooling by CMB photons. Secondly, in Section
\ref{sec:ionization} we consider the (re)ionization of hydrogen
in the \revision{collisional ionization} case, if the background gas is already
partially neutral. We assume that hydrogen gets collisionally
ionized once the gas reaches a characteristic temperature
$T_{\rm ion} \sim 10^4$ K, and that this ionization proceeds
roughly at constant temperature. Thirdly, in Section
\ref{sec:inner} we study the innermost region where the gas is
fully ionized and adiabatically compressed. We account for the
change of the adiabatic index once electrons become
relativistic. The final outcome of this calculation is the gas
temperature near the event horizon, which, alongside the
accretion rate, determines the luminosity of the accreting BH,
as we shall see in Section \ref{sec:luminosity}. Figure
\ref{fig:T_of_r} illustrates the temperature profile in the
various regions considered. We conclude this Section by
considering the effect of PBH velocities.

\subsection{Outermost, constant-ionization-fraction region} \label{sec:outer}

\subsubsection{Setup}

We consider spherical accretion of a pure hydrogen\footnote{Accounting for helium is conceptually straightforward but would add unneeded complications for the order-of-magnitude calculation presented here.} gas onto an isolated point mass $M$, bathed in the quasi-uniform CMB radiation field (we check the validity of the isolated-PBH assumption in Appendix \ref{app:isolated}). In general, one should solve for the time-dependent fluid, heat and ionization equations, all of which are coupled. For simplicity we shall assume a constant ionization fraction \revision{$x_e = \overline{x}_e$ in the outermost region, equal to the background value}. As long as the characteristic accretion timescale is much shorter than the Hubble timescale, one can make the steady-state approximation. Ref.~\cite{Ricotti_07} showed that this is the case for $M \lesssim 3 \times 10^4\, M_{\odot}$, so we shall limit ourselves to this mass range. In this outermost region, far from the BH horizon, a Newtonian treatment is very accurate.

We denote by $v  \equiv v_r < 0$ the peculiar radial velocity (i.e.~the velocity with respect to the Hubble flow) of the accreted gas. The steady-state mass and momentum equations for the fluid are 
\barr
4 \pi r^2 \rho |v| &=& \dot{M} = \textrm{const},\\
v \frac{d v}{d r} &=& - \frac{G M}{r^2}  - \frac1{\rho} \frac{d P}{dr} - \frac43 \frac{\overline{x}_e \sigma_{\rm T} \rho_{\rm cmb}}{m_p c} v, \label{eq:momentum}
\earr
where the pressure $P$ is
\beq
P = \frac{\rho}{m_p}(1 + \overline{x}_e) T,
\eeq
and the last term in the momentum equation is the drag force due to Compton scattering of the ambient nearly homogeneous CMB photons with energy density $\rho_{\rm cmb}$ \cite{Ricotti_07}, $\sigma_{\rm T}$ being the Thomson cross section. Note that we have neglected the self-gravity of the accreted gas, which is valid for $M \lesssim 3 \times 10^5\, M_{\odot}$ \cite{Ricotti_07}. \revision{Consistent with our steady-state approximation, we also neglected the Hubble drag term $H v$, which is of the same order as the neglected partial time derivative $\partial v/\partial t$.}

The fluid equation must be complemented by the heat equation. For simplicity we shall only consider Compton cooling by CMB photons \cite{Peebles_68} as a heat sink in this region. The steady-state heat equation is then
\beq
v \rho^{2/3} \frac{d}{dr}\left( \frac{T}{\rho^{2/3}}\right) = \frac{8 \overline{x}_e \sigma_{\rm T}\rho_{\rm cmb}}{3 m_e c (1 + \overline{x}_e)}(T_{\rm cmb} - T), \label{eq:heat-compt}
\eeq
where $T_{\rm cmb}$ is the temperature of CMB photons. Since we only consider PBH masses for which the accretion timescale is shorter than the Hubble timescale, whenever Compton cooling becomes relevant to the accretion problem, it is even more important for the background temperature evolution, and enforces $T_{\infty} = T_{\rm cmb}$.

If Compton drag and cooling were negligible, one would recover the the classic Bondi accretion problem \cite{Bondi_52}, the characteristic velocity, length and timescales of which are
\barr
v_{\rm B} \equiv \sqrt{P_{\infty}/\rho_{\infty}},\  \ \ r_{\rm B} \equiv \frac{G M}{v_B^2}, \ \ \ t_{\rm B} \equiv \frac{G M}{v_B^3}, \label{eq:rB-tB}
\earr
where $P_{\infty}$ and $\rho_{\infty}$ are the gas pressure and density far from the point mass ($\rho_{\infty} = \overline{\rho}_b$, the mean baryon density).

\begin{figure}[h]
\includegraphics[width = \columnwidth]{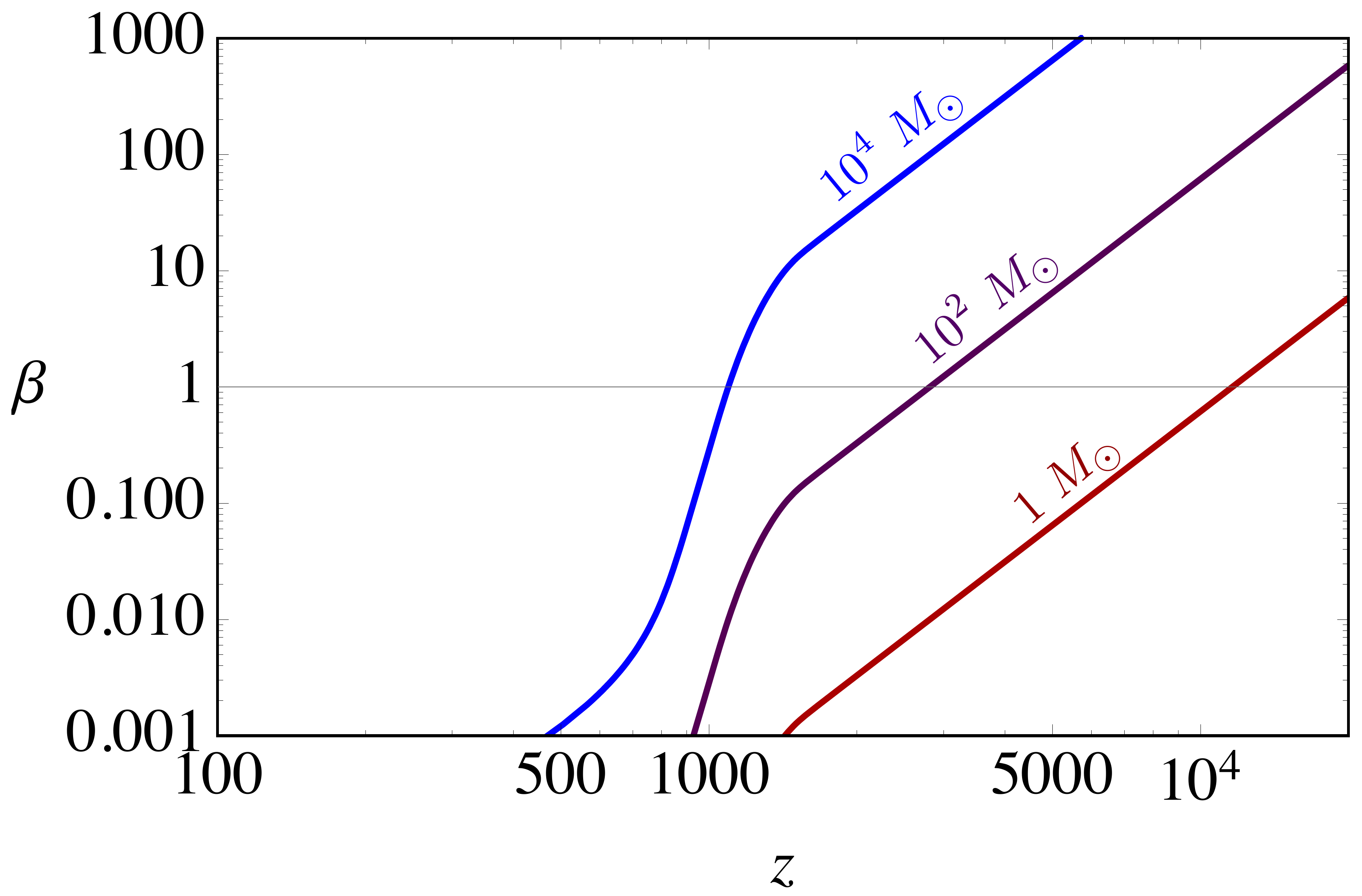}
\includegraphics[width = \columnwidth]{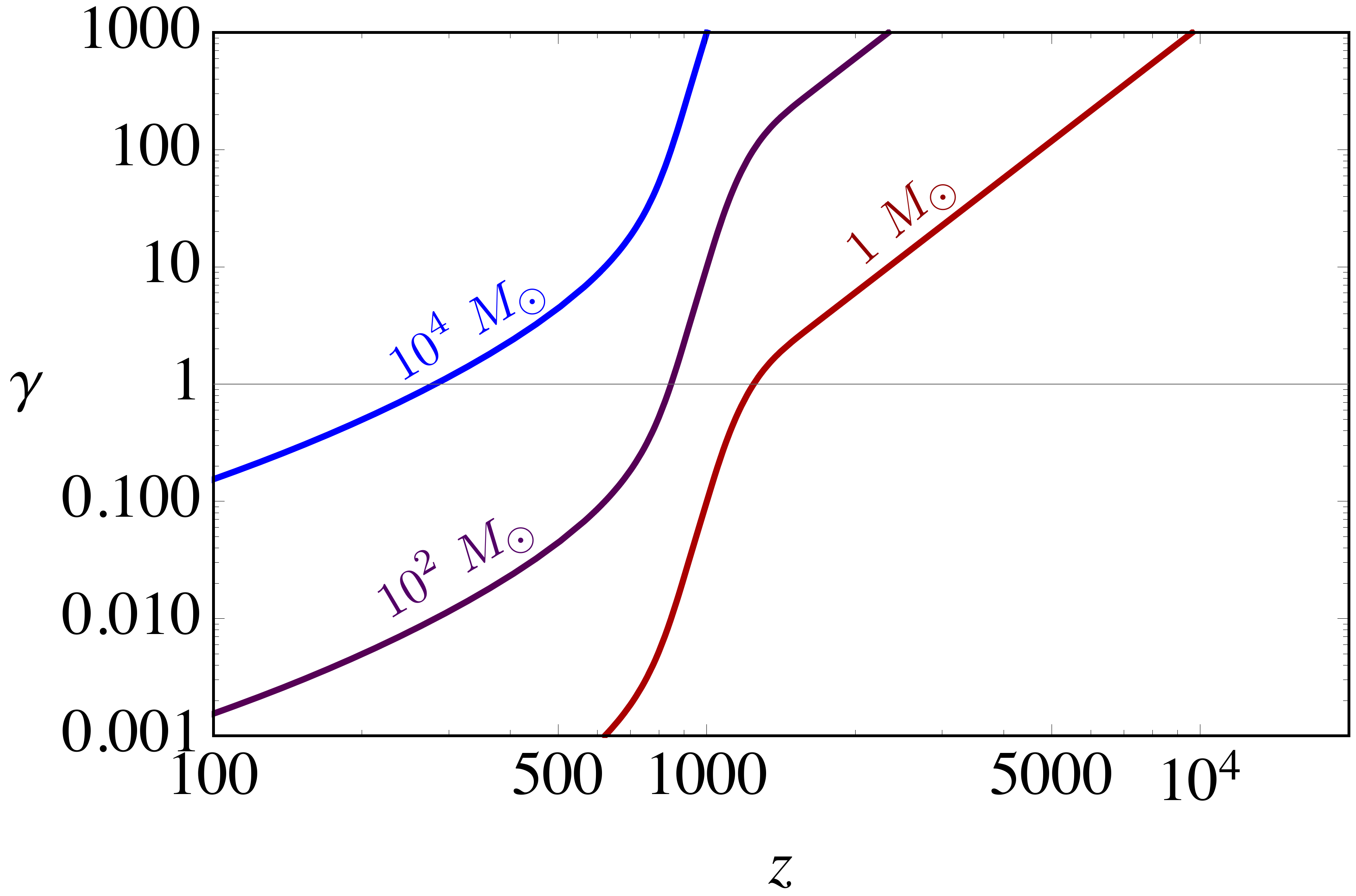}
\caption{Characteristic dimensionless Compton drag rate $\beta$ [Eq.~\eqref{eq:beta-def}, \emph{upper panel}] and Compton cooling rate $\gamma$ [Eq.~\eqref{eq:gamma-def}, \emph{lower panel}], as a function of redshift, and for PBH masses $M = 1, 10^2$ and $10^4\, M_{\odot}$, from bottom to top. \revision{Both are evaluated for a standard recombination and thermal history, with the substitution $v_{\rm B} \rightarrow v_{\rm eff}$ as described in Section \ref{sec:velocities}.}} 
\label{fig:beta_gamma}
\end{figure}

It is best to rewrite the problem in terms of dimensionless variables $x  \equiv r/r_{\rm B}$, $u \equiv v/v_{\rm B}$, $\hat{\rho} \equiv \rho/\rho_{\infty}$, $\hat{T} \equiv T/T_{\infty}$. We also define the dimensionless constants
\barr
\lambda &\equiv& \frac{\dot{M}}{4 \pi \rho_{\infty} r_{\rm B}^2 v_{\rm B}}, \\
\beta &\equiv&  \frac43 \frac{\overline{x}_e \sigma_{\rm T}\rho_{\rm cmb} }{m_p c} t_{\rm B}, \label{eq:beta-def}\\
\gamma &\equiv& \frac{8 \overline{x}_e \sigma_{\rm T}\rho_{\rm cmb} }{3 m_e c (1 + \overline{x}_e)} t_{\rm B}= \frac{2 m_p}{m_e (1 + \overline{x}_e)} \beta \gg \beta. \label{eq:gamma-def}
\earr
\change{We show in Fig.~\ref{fig:beta_gamma} the dimensionless Compton drag and cooling rates $\beta$ and $\gamma$, as a function of redshift and PBH mass.} In terms of these variables the problem to solve is 
\barr
\hat{\rho} x^2 |u| &=& \lambda, \label{eq:rho u x2}\\
u \frac{du}{dx} &=& - \frac1{x^2} - \frac{1}{\hat{\rho}} \frac{d}{dx}(\hat{\rho} \hat{T}) - \beta u, \label{eq:momentum-dimensionless}\\
u \hat{\rho}^{2/3} \frac{d}{dx}\left( \frac{\hat{T}}{\hat{\rho}^{2/3}}\right) &=& \gamma(1 - \hat{T}), \label{eq:Temp-dimensionless}
\earr
with asymptotic conditions $\hat{\rho} \rightarrow 1$ and $\hat{T} \rightarrow 1$ at $x \rightarrow \infty$. 

Before moving on, let us note that the PBH mass does not grow
significantly in a Hubble time \cite{Carr:1974nx}. Indeed,
\barr
\frac{\dot{M}}{H M} = \frac{4 \pi \lambda \overline{\rho}_b (G M)^2}{H M v_{\rm B}^3} = \frac{4 \pi \lambda G \overline{\rho}_b t_B}{H}= \frac32 \lambda \frac{\overline{\rho}_b}{\overline{\rho}_{\rm tot}} H t_{\rm B} ,~~~\label{eq:Mdot/HM}
\earr
where in the last equality we used Friedman's equation for the Hubble rate $H$. Therefore, provided the steady-state approximation is valid (i.e. $t_{\rm B} \ll H^{-1}$), we see that $\dot{M}\ll H M$.

\subsubsection{Solution for $\beta \ll \gamma \ll 1$}

When both Compton drag and cooling are negligible, we recover the classic Bondi problem \cite{Bondi_52} for an adiabatic gas, with $\hat{T} = \hat{\rho}^{2/3}$. In this case the momentum equation can be rewritten as a conservation equation
\beq
\frac12 u^2 - \frac1{x} + \frac52 (\hat{\rho}^{2/3} - 1) = 0. \label{eq:mom-cons-ad}
\eeq
Using Eq.~\eqref{eq:rho u x2} and multiplying by $2 x$, we get
\beq
x u^2  + \frac{5 \lambda^{2/3}}{(x u^2)^{1/3}} = 5 x + 2.
\eeq
The left-hand-side reaches a minimum at $x u^2 = (5/3)^{3/4} \lambda^{1/2}$, with value $4(5/3)^{3/4} \lambda^{1/2}$. For a solution to exist for all $x$, this has to be less than 2, the minimum of the right-hand-side, implying\footnote{The difference of our maximum value of $\lambda$ and the usually quoted value of 1/4 comes from our normalization of velocities with $v_{\rm B}$ rather than the adiabatic sound speed at infinity, which is $(5/3)^{1/2} v_{\rm B}$.} $\lambda \leq \lambda_{\rm ad} \equiv \frac14 (3/5)^{3/2}$. Though all solutions with sub-critical $\lambda \leq \lambda_{\rm ad}$ are a priori acceptable\footnote{This is not the case for the Bondi problem with adiabatic index $< 5/3$, for which sub-critical solutions have a velocity that tends to zero near the origin, which is unphysical.}, we shall assume, like Bondi, that the physically realized solution is that of maximum accretion, i.e. that 
\beq
\lambda =  \lambda_{\rm ad} \equiv \frac14 \left(\frac35\right)^{3/2} \approx 0.12.
\eeq 
Combining Eqs.~\eqref{eq:mom-cons-ad} and \eqref{eq:rho u x2} one can show that the asymptotic behaviors of fluid variables for $x \ll 1$ are
\barr
u(x) &\approx& - \frac1{\sqrt{2}} x^{-1/2},\\
\hat{\rho}(x) &\approx& \left(\frac3{10}\right)^{3/2} x^{-3/2},\\
\hat{T}(x) &\approx& \frac{3}{10} x^{-1}.  \label{eq:T-asym-ad}
\earr

\subsubsection{Solution for $\beta \ll 1$ and $\gamma \gg 1$} \label{sec:compt_cool}

If $\gamma \gg1$ Compton cooling efficiently maintains $\hat{T} \approx 1$ down to $x \sim \gamma^{-2/3} \ll 1$. At that point pressure forces are negligible relative to gravity, and the temperature is no longer relevant to the other fluid variables. We may therefore first solve the isothermal Bondi problem for the fluid variables, and deduce the temperature profile from them. For the isothermal Bondi problem with $\hat{T} = 1$, the conserved quantity is now
\beq
\frac12 u^2 - \frac1x + \ln(\hat{\rho}) = 0.
\eeq
Here again one can show that there exists a maximum value of $\lambda$ for which the problem has a solution. For sub-critical $\lambda$, however, the velocity tends to zero towards the origin and the density diverges unphysically as $\rme^{1/x}$. Therefore the physically valid solution is that with the critical accretion rate
\beq
\lambda = \lambda_{\rm iso} \equiv \frac14 \rme^{3/2} \approx 1.12.
\eeq
It is sensible that the accretion rate is larger in the isothermal case than in the adiabatic case. \change{Indeed, the temperature is larger in the adiabatic case, providing a larger pressure support counterbalancing gravity.} For $x \ll 1$ the velocity reaches the free-fall solution $u(x) \approx - \sqrt{2/x}$ and the density is then $\hat{\rho}(x) \propto  x^{-3/2}$. Inserting these asymptotic forms into the heat equation, we get
\beq
\frac{\sqrt{2}}{x^{3/2}} \frac{d}{dx}(x \hat{T}) = \gamma(\hat{T}-1).
\eeq
One can write an explicit integral solution to this equation. In particular we find the asymptotic limit for $x \ll \gamma^{-2/3}$,
\beq
\hat{T}(x) \approx \left(\frac{4}{3}\right)^{1/3}  \frac{\Gamma(2/3) }{\gamma^{2/3}x} \approx \frac{1.5}{\gamma^{2/3} x}, \label{eq:T-asym-iso}
\eeq
where $\Gamma$ is Euler's Gamma function.

\subsubsection{Solution for $\beta \ll 1$ and arbitrary $\gamma$}

For arbitrary values of $\gamma$ (while $\beta \ll 1$) the momentum equation can no longer be rewritten as a conservation equation and one must solve explicitly the coupled fluid and heat equations, and determine the accretion ``eigenvalue" $\lambda$ numerically. 

\begin{figure}
\includegraphics[width = \columnwidth]{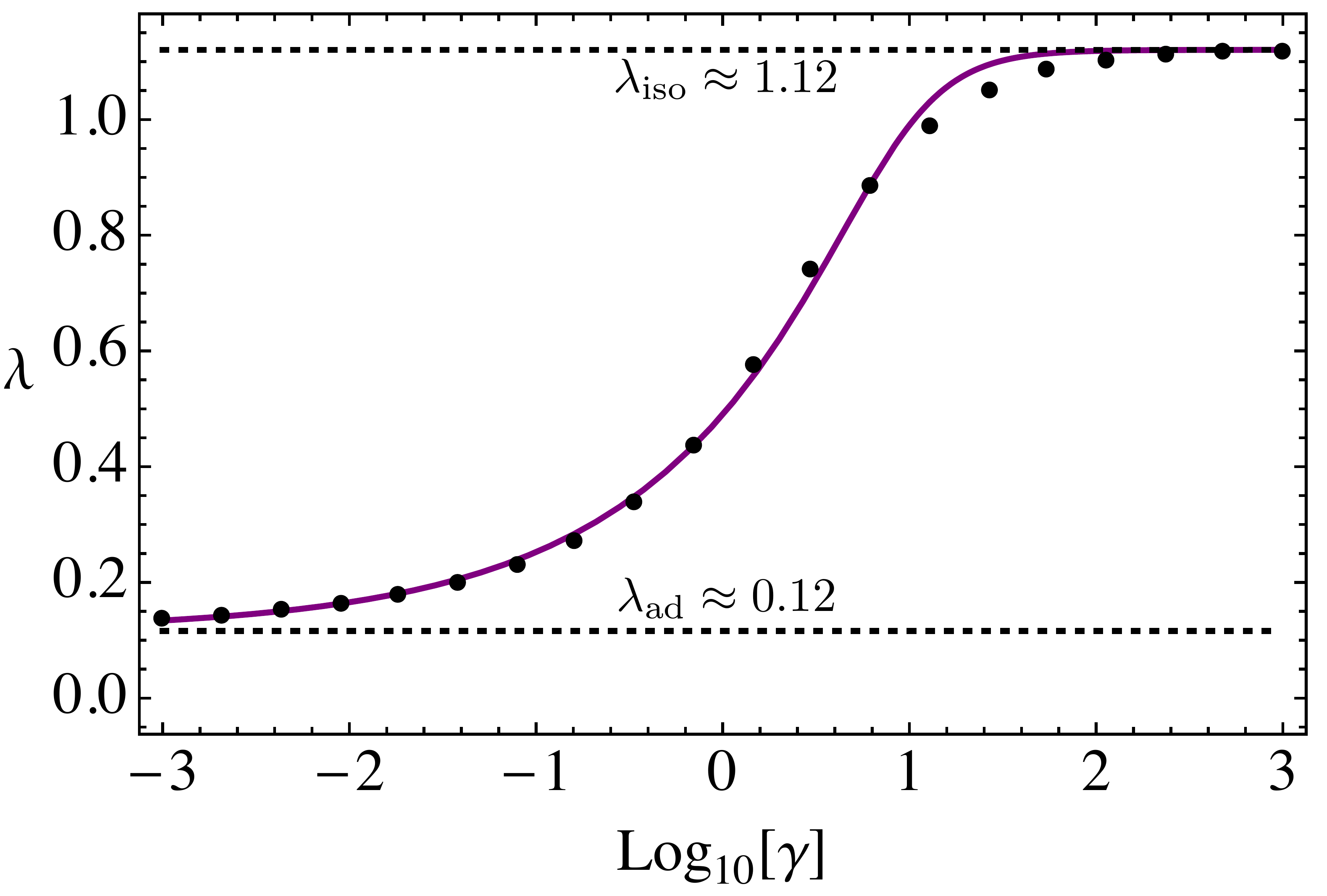}
\caption{Dimensionless accretion rate $\lambda$ as a function of the dimensionless Compton cooling rate $\gamma$. Black circles are our numerical results and the purple line is our analytic fit, Eq.~\eqref{eq:lambda_fit}.} \label{fig:lambda_gamma}
\end{figure}

We re-write the system \eqref{eq:momentum-dimensionless}-\eqref{eq:Temp-dimensionless} in the form
\barr
\left(u - \frac53 \frac{\hat{T}}{u}\right) \frac{du}{dx} &=& \frac{10}3 \frac{\hat{T}}{x} - \frac1{x^2} - \gamma \frac{1-\hat{T}}{u}, \label{eq:u}\\
\frac{d\hat{T}}{dx}  &=& - \frac23 \frac{\hat{T}}{u} \frac{d u}{dx} - \frac43 \frac{\hat{T}}{x} + \gamma \frac{1-\hat{T}}{u}.\label{eq:That}
\earr
The boundary conditions at large radii are $u(x) = -\lambda/x^2, \hat{T}(x) = 1$. We see that the system is singular \change{at the point $x_*$ where the velocity reaches the local adiabatic sound speed,}
\beq
u_* =-\sqrt{ 5 \hat{T}_*/3}, \label{eq:u*}
\eeq
unless this condition is met simultaneously with
\beq
\frac{10}3 \frac{\hat{T}_*}{x_*} - \frac1{x_*^2} - \gamma \frac{1-\hat{T}_*}{u_*} = 0, \label{eq:T*}
\eeq
so that the right-hand-side of Eq.~\eqref{eq:u} vanishes, leading to a finite derivative. There is a single value $\lambda_*$ for which these two conditions are satisfied simultaneously: larger $\lambda$ lead to a singularity while for lower values $du/dx$ changes sign before the singularity, and the velocity unphysically tends to zero at the origin. We find $\lambda_*$ by bisection: starting with $\lambda_{\rm min} \equiv 0 < \lambda_* < \lambda_{\rm max} \equiv 2$, we set $\lambda = (\lambda_{\rm min} + \lambda_{\max})/2$ and integrate the system numerically from $x = 100$ towards $x=0$, until either the singularity or $du/dx = 0$ is reached. In the former case, we set $\lambda_{\max} = \lambda$ at the next step, and in the latter case, we set $\lambda_{\min} = \lambda$, so that $\lambda_{\min} < \lambda_* < \lambda_{\max}$ at each step. We do so until the fractional difference between $\lambda_{\max}$ and $\lambda_{\min}$ is less than a small error tolerance, typically $10^{-6}$. We show the resulting function $\lambda(\gamma)$ in Fig.~\ref{fig:lambda_gamma}. We find that the following analytic expression is a good fit to the numerical results:
\beq
\lambda(\gamma; \beta \ll 1) \approx  \lambda_{\rm ad} + (\lambda_{\rm iso} - \lambda_{\rm ad}) \left(\frac{\gamma^2}{88 + \gamma^2}\right)^{0.22}.\label{eq:lambda_fit}
\eeq

While it is relatively simple to obtain a very precise value of $\lambda$ numerically, obtaining a precise asymptotic limit of $\hat{T}(x)$ at $x \rightarrow 0$ proved to be more challenging. Keeping in mind that this calculation is an order-of-magnitude estimate, we simply assume the following expression, interpolating between the adiabatic case \eqref{eq:T-asym-ad} and the quasi-isothermal case \eqref{eq:T-asym-iso}:
\beq
\tau \equiv \underset{x \rightarrow 0}{\lim}(x \hat{T}) \approx \frac{1.5}{5 + \gamma^{2/3}}. \label{eq:tau_fit}
\eeq
Inserting $T \approx \tau/x$, $u \approx - \omega/x^{1/2}$, $\rho \propto  x^{-3/2}$ in the momentum equation \eqref{eq:momentum-dimensionless}, we find $\omega = \sqrt{2 - 5 \tau}$. In summary, the asymptotic values of the temperature, velocity and density fields are 
\barr
\hat{T}(x) &\approx& \frac{\tau}{x}, \label{eq:T-largebeta}\\
u(x) &\approx& -\sqrt{\frac{2 - 5 \tau}{x}}, \label{eq:u-largebeta}\\
\hat{\rho}(x) &\approx& \frac{\lambda}{\sqrt{2 - 5 \tau}} x^{-3/2}.\label{eq:rho-largebeta}
\earr

\subsubsection{Solution for $1 \lesssim \beta \ll  \gamma$}

\change{When Compton drag is significant ($\beta \gtrsim 1$),} there is no longer any conserved quantity, even in the quasi-isothermal case. We can simply determine the asymptotic value of $\lambda$ for $\beta \gg 1$ by considering the momentum equation at $x \ll 1$, where the pressure force is negligible with respect to gravity. In this regime we find $u \approx - 1/(\beta x^2)$, implying that $\lambda \rightarrow \beta^{-1}$ for large $\beta$. Physically, the drag force balances the gravitational force, i.e.~the velocity reaches the terminal velocity. Once $x \lesssim \beta^{-2/3} \gg \gamma^{-2/3}$, the advection term $u (du/dx)$ becomes dominant over the drag term $-\beta u$ and the velocity reaches the free-fall solution $u \approx -\sqrt{2/x}$. Since this occurs at a radius much larger than $\gamma^{-2/3}$, the asymptotic behavior or $\hat{T}$, is still given by Eqs.~\eqref{eq:T-largebeta} and \eqref{eq:tau_fit}. The effect of Compton drag is therefore only to change the accretion rate. 

Ref.~\cite{Ricotti_07} find the following analytic approximation for $\lambda(\beta)$, valid for all values of $\beta$ (but for $\gamma \gg 1$ only, as they consider isothermal accretion):
\beq
\lambda(\gamma \gg 1; \beta) \approx \exp \left[ \frac{9/2}{3 + \beta^{3/4}}\right] \frac1{(\sqrt{1 + \beta}+1)^2}.
\eeq
For general $\gamma$ and $\beta$ we may use the following approximation for the dimensionless accretion rate:
\beq
\lambda(\gamma, \beta)  = \frac{\lambda(\gamma; \beta \ll 1) \lambda(\gamma \gg 1; \beta)}{\lambda_{\rm iso}}. \label{eq:lambda-full}
\eeq
This approximation is well justified since $\beta  \ll \gamma$. As a consequence, either $\beta \ll 1$ or $\gamma \gg 1$. 

\begin{figure}
\includegraphics[width = \columnwidth]{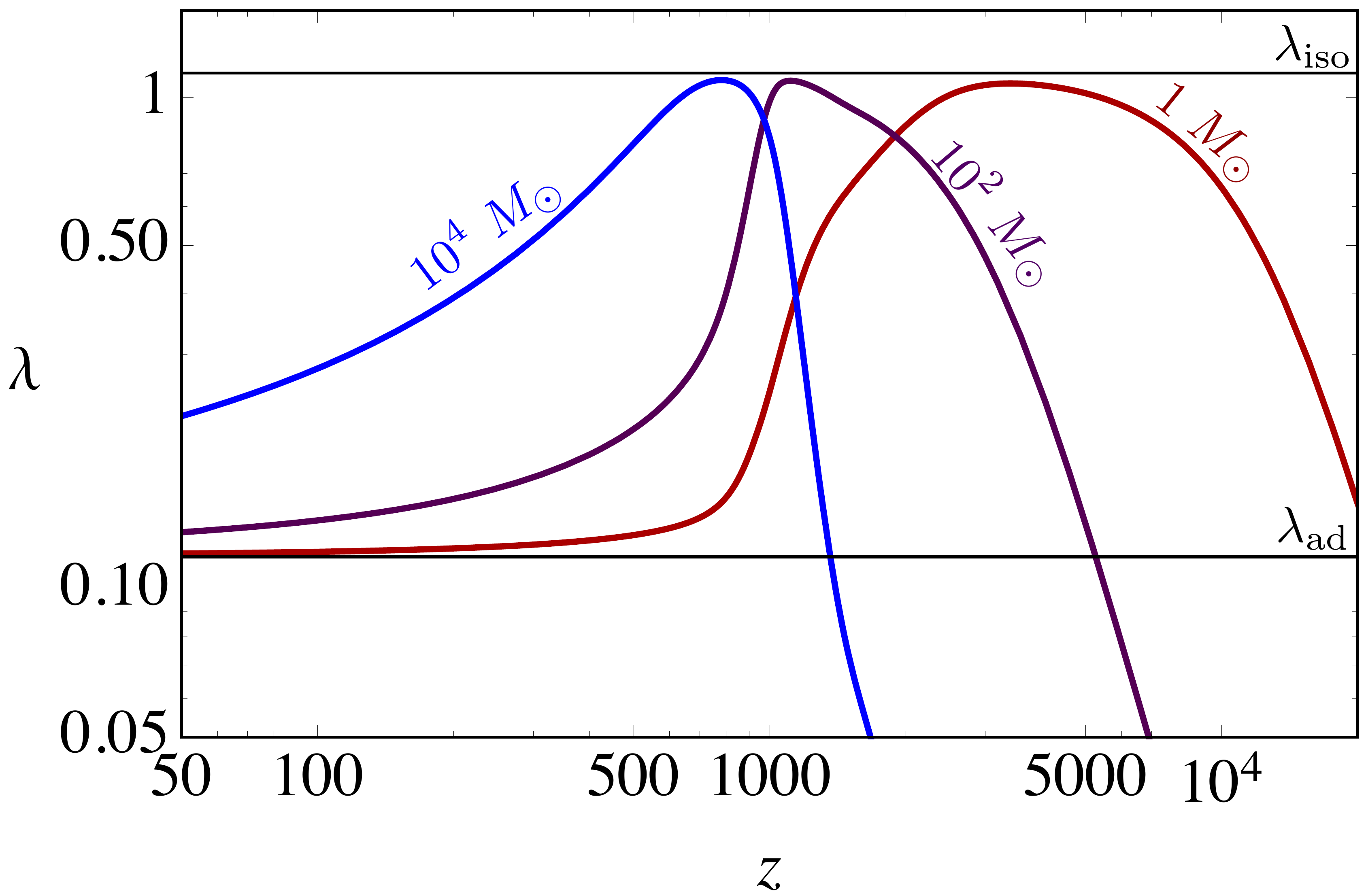}
\includegraphics[width = \columnwidth]{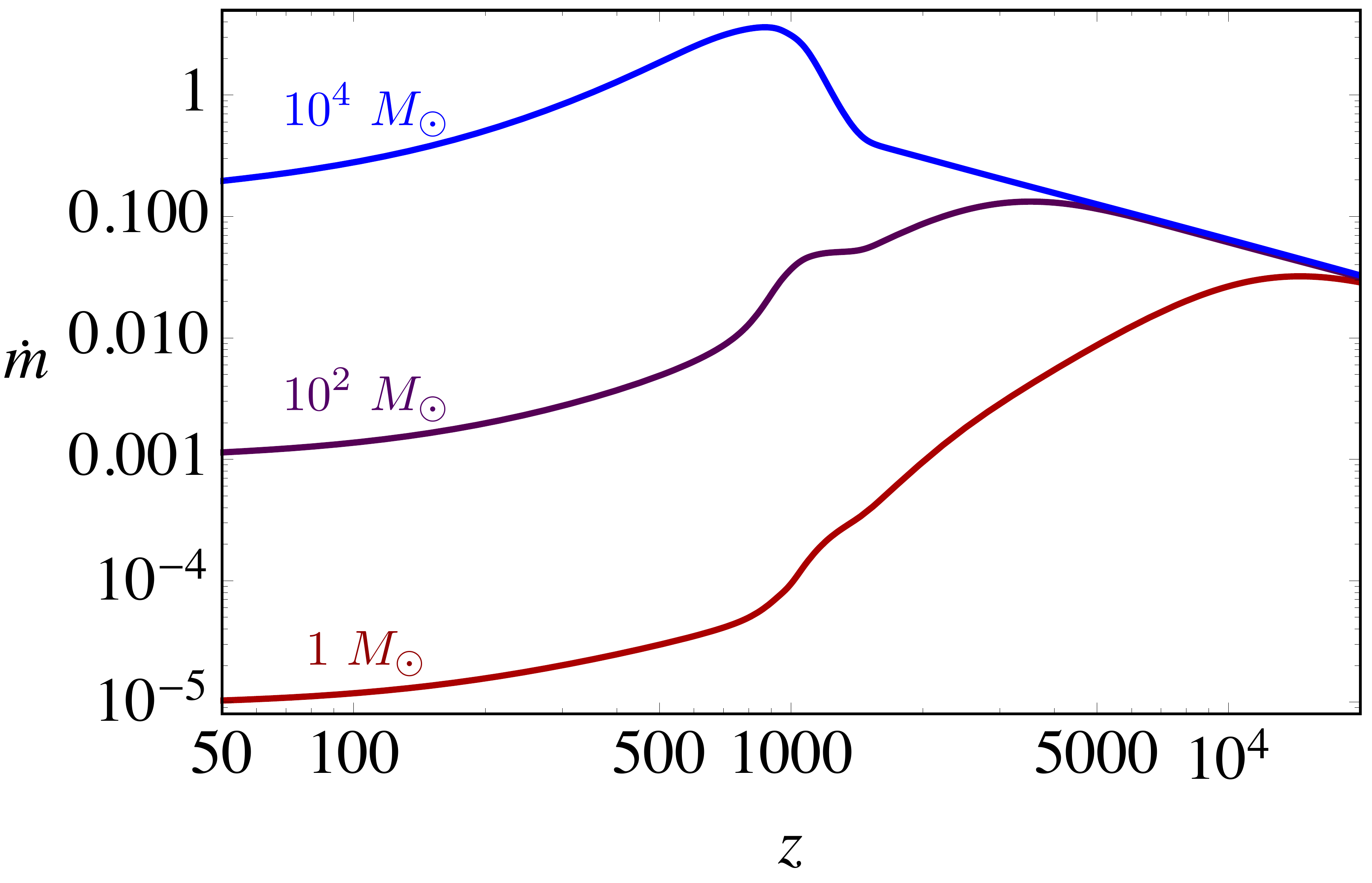}
\caption{Characteristic dimensionless accretion rate $\lambda$ (\emph{upper panel}) and accretion rate normalized to the Eddington value $\dot{m} \equiv \dot{M} c^2/L_{\rm Edd}$ (\emph{lower panel}) as a  function of redshift, for PBH masses $1, 10^2$ and $10^4\, M_{\odot}$. These quantities are evaluated with substitution $v_{\rm B} \rightarrow v_{\rm eff}$ as described in Section \ref{sec:velocities}. } \label{fig:lambda}
\end{figure}

The dimensionless accretion rate $\lambda$ is the first main result of this Section. We show its evolution as a function of redshift for several PBH masses in Fig.~\ref{fig:lambda}. While ROM do account for Compton drag following the analysis of Ref.~\cite{Ricotti_07}, they implicitly assume that $\gamma \gg 1$ at all times. In other words, they do not account for the factor of $\sim 10$ decrease of $\lambda$ at low redshift when Compton cooling becomes negligible and the accretion becomes mostly adiabatic. \change{Figure \ref{fig:lambda} also shows the evolution of the accretion rate normalized to the Eddington rate, $\dot{m} \equiv \dot{M} c^2/L_{\rm Edd}$.} 

\subsection{\revision{Collisional ionization region}} \label{sec:ionization}

\revision{If the emerging radiation field is too weak to photoionize the gas, it eventually gets collisionally ionized as it is compressed and heated up}. We assume that this proceeds roughly at constant temperature $T \approx T_{\rm ion}\approx 1.5 \times 10^4$. Indeed, if ionization proceeds through collisional ionizations balanced by radiative recombinations, the equilibrium ionization fraction only depends on temperature, with a sharp transition at $ T \approx 1.5 \times 10^4$ K (for instance, using Eq.~(2) or Ref.~\cite{Nobili_91}, we get $x_e = (0.01, 0.5, 0.99)$ at $T = (1.1, 1.5, 2.5) \times 10^4$ K, respectively). 

Getting back to dimensionful variables, we found in the previous section that at small radii, 
\barr
T(r) \approx \tau T_{\infty} \frac{r_{\rm B}}{r}, \label{eq:T-asympt}
\earr
where $\tau$ is a dimensionless constant at most equal to 3/10, and smaller when Compton cooling is important. The effect of the ionization region is only relevant once the global free-electron fraction $\overline{x}_e$ falls significantly below unity, i.e. for $T_{\infty} \lesssim 3000$ K $\ll T_{\rm ion}$. Therefore we expect the ionization region to be reached deep inside the Bondi radius, where the asymptotic behavior \eqref{eq:T-asympt} is accurate. The ionization region therefore starts at radius
\beq
r_{\rm ion}^{\rm start} \approx \tau \frac{T_{\infty}}{T_{\rm ion}} r_{\rm B}, \label{eq:r_ion_start}
\eeq
where the density is, from Eq.~\eqref{eq:rho-largebeta}
\beq
\rho_{\rm ion}^{\rm start} \approx  \frac{\lambda}{\sqrt{2 - 5 \tau}} \rho_{\infty}\left(\frac{T_{\rm ion}}{\tau T_{\infty}}\right)^{3/2}. \label{eq:rho_ion_start}
\eeq
We assume that ionization proceeds through collisions of neutral hydrogen atoms with free electrons. This process only redistributes the internal energy of the gas, i.e.~does not generate any net heat. However it does lead to a temperature decrease as the number of free particles increase and some of their energy is used to ionize the gas. If the temperature is to remain constant through the ionization region, this effect must be compensated by the temperature increase due to the adiabatic compression of the gas. In equations, we write the first law of thermodynamics:
\barr
\Delta \left(\frac32 (1+ x_e) T - (1 - x_e) E_{\rm I}\right) = - (1 + x_e) T \rho \Delta(1/\rho),~~~~~
\earr
where the second term in the internal energy \change{on the left-hand-side} accounts for the binding energy $E_{\rm I} = 13.6$ eV of neutral hydrogen atoms. Assuming the temperature remains constant throughout the ionization region we arrive at the simple relation between changes in density and ionization fraction:
\beq
\Delta \ln \rho =  \left(\frac32 + \frac{E_{\rm I}}{T_{\rm ion}}\right) \Delta \ln (1 + x_e).
\eeq
Therefore the ratio of the density at the end of the ionization region to that at its beginning is
\beq
\frac{\rho_{\rm ion}^{\rm end}}{\rho_{\rm ion}^{\rm start}} =  \left(\frac{2}{1 + \overline{x}_e}\right)^{\frac32 + \frac{E_{\rm I}}{T_{\rm ion}}} \approx \left(\frac{2}{1 + \overline{x}_e}\right)^{12}, \label{eq:chi}
\eeq
where we took $T_{\rm ion} \approx 1.5 \times 10^4$ K. Assuming that $\rho \propto r^{-3/2}$ throughout the region, we get
\beq
\frac{r_{\rm ion}^{\rm end}}{r_{\rm ion}^{\rm start}} \approx \left(\frac{1 + \overline{x}_e}2\right)^{8} \label{eq:r_ion_ratio}.
\eeq
We see that the ionization region may extend by a factor of $\sim 300$ in radius if $\overline{x}_e \ll 1$. This is consistent with Shapiro's results \cite{Shapiro_73}, who finds an ionization region extending over a factor $\sim 10^3$ in radius for accretion from a neutral gas. Note that we have neglected the heat loss due to collisional excitations followed by radiative decays, as they cannot be simply included in our basic treatment. We have also neglected Compton cooling by CMB photons, which may become relevant again once the ionization fraction increases. Accounting for these cooling mechanisms would imply a larger density contrast $\rho^{\rm end}/\rho^{\rm start}$ hence a more extended ionization region, and an overall larger suppression of the temperature near the PBH horizon.

\revision{If the radiation field from the accreting PBH is intense enough, it may photoionize the gas beyond $r_{\rm ion}$, in which case there is no collisional ionization region. To group both cases we define $\rho_{\rm ion}^{\rm end} =\rho_{\rm ion}^{\rm start}$ in that case, so that in general }
\barr
\frac{\rho_{\rm ion}^{\rm end}}{\rho_{\rm ion}^{\rm start}} = \chi \equiv 
\begin{cases}
\left(\frac{2}{1 + \overline{x}_e}\right)^8   &\textrm{(collisional ionization)}, \\
1   &\textrm{(photoionization)}. \label{eq:chi-general}
\end{cases}
\earr

\subsection{Innermost adiabatic region}\label{sec:inner}

Once the gas is fully ionized, it resumes adiabatic compression (we justify in Appendix \ref{sec:free-free cooling} that free-free cooling can neglected for the mass range considered). The thermal energy density of the ionized plasma is 
\beq
u = \frac32  n_e \left(1 + f(T/m_e c^2)\right) T,
\eeq
where the dimensionless function $f$ accounts for the fact that electrons are potentially relativistic, and has asymptotic limits $f(X \ll 1) = 1$ and $f(X \gg 1) = 2$. The pressure remains unchanged $P = 2n_e T$, and the first law of thermodynamics can then be written
\beq
\frac32 \left[1 + f(X) + X f'(X)\right] \frac{dT}{T} = 2 \frac{d \rho}{\rho}, \ \ \ \ X \equiv \frac{T}{m_e c^2}.
\eeq
This can be integrated to give
\beq
\frac{\rho_2}{\rho_1} = \frac34 \int_{X_1}^{X_2}  \left[1 + f(X) + X f'(X)\right] \frac{dX}{X}. \label{eq:rho2_ov_rho1}
\eeq
We have computed the function $f$ explicitly and find that it is well approximated by the simple functional form
\beq
f(X) \approx 1 + \frac{X}{X+0.73}.  \label{eq:f-approx}
\eeq
With this simple analytic form Eq.~\eqref{eq:rho2_ov_rho1} can be integrated analytically to obtain $\rho_2(T_2)$. We invert this relation numerically and obtain the following approximation, valid for $T_1 \ll m_e c^2$ and arbitrary $T_2$:
\barr
\frac{T_2}{m_e c^2} &\approx& \mathcal{F}\left(\frac{T_1}{m_e c^2} \left(\frac{\rho_2}{\rho_1}\right)^{2/3}\right),\\
\mathcal{F}(Y) &\equiv& Y \left(1 + \frac{Y}{0.27}\right)^{-1/3}. \label{eq:T-of-rho}
\earr
This recovers the expected asymptotic behaviors $T \propto \rho^{2/3}$ for $T \lesssim m_e c^2$ and $T \propto \rho^{4/9}$ for $T \gtrsim m_e c^2$ \change{and moreover gives an accurate result for arbitrary temperatures.}

We may now finally compute the gas temperature near the Schwarzschild radius $r_{\rm S}$. The velocity there nears the speed of light, $|v| \approx c$, so the density is 
\barr
\rho_{\rm S} &=& \frac{\lambda}{(c/v_{\rm B}) (r_{\rm S}/r_{\rm B})^2} \rho_{\infty}= \frac{\lambda}{4 (v_{\rm B}/c)^3} \rho_{\infty} \nonumber\\
&=& \frac{\lambda}{4} \left(\frac{m_p c^2}{(1+ \overline{x}_e) T_{\infty}}\right)^{3/2} \rho_{\infty}.
\earr
At the end of the ionization region, the temperature is $T_{\rm ion}$ and, from Eqs.~\eqref{eq:rho_ion_start} and \eqref{eq:chi-general}, \revision{the density is} 
\beq
\rho_{\rm ion}^{\rm end} = \chi \frac{\lambda}{\sqrt{2 - 5 \tau}} \left(\frac{T_{\rm ion}}{\tau T_{\infty}}\right)^{3/2} \rho_{\infty},
\eeq
Using Eq.~\eqref{eq:T-of-rho} with $T_1 = T_{\rm ion}$ and $\rho_1 = \rho_{\rm ion}^{\rm end}$, we finally obtain the temperature $T_{\rm S}$ at the Schwarzschild radius: 
\beq
T_{\rm S} = m_e c^2 \mathcal{F}(Y_{\rm S}),
\eeq
where $\mathcal{F}$ is given by Eq.~\eqref{eq:T-of-rho} and
\barr
Y_{\rm S} &\equiv& \frac{T_{\rm ion}}{m_e c^2} \left(\frac{\rho_{\rm S}}{\rho_{\rm ion}^{\rm end}}\right)^{2/3}\nonumber\\
&=& \chi^{-2/3} \left(\frac2{1 + \overline{x}_e}\right)  \frac{\tau}{4} \left(1 - \frac52 \tau\right)^{1/3}\frac{m_p}{m_e}. \label{eq:Y-final}
\earr
It is interesting to compare this result to those of Shapiro \cite{Shapiro_73}, who did not consider Compton cooling (i.e. $\tau = 3/10$), \revision{assumed that photoionizations are negligible (in which case we have $\chi^{-2/3} \approx [(1+ \overline{x}_e)/2]^8$), and only studied the cases $\overline{x}_e = 1$ or 0.} In the former case, we find 
\beq
Y_{\rm S} \approx \frac3{40} 4^{-1/3} \frac{m_p}{m_e} \approx 10^2 \gg 1 
\eeq
and as a result electrons are relativistic at the Schwarzschild radius, with temperature
\barr
T_{\rm S} &\approx& m_e c^2 0.27^{1/3} Y_{\rm S}^{2/3} \approx 0.08 (m_p c^2)^{2/3} (m_e c^2)^{1/3} \nonumber\\
&\approx& 0.7 \times 10^{11} \textrm{K},
\earr
in excellent agreement with Shapiro's result (see also \cite{ST_83}). In the case of a neutral background, taking $T_{\rm ion} = 1.5 \times 10^4$ K, $Y_{\rm S}$ is a factor $\sim 2^{-7}$ smaller, i.e. $Y_{\rm S} \approx 0.7$, so electrons are marginally relativistic at the horizon, with $T_{\rm S} \approx 0.4~ m_e c^2 \approx 2.5 \times 10^9$ K. This is a factor of $\sim 2$ higher than Shapiro's result, consistent with our neglect of collisional excitations in the ionization region.

\begin{figure}
\includegraphics[width = \columnwidth]{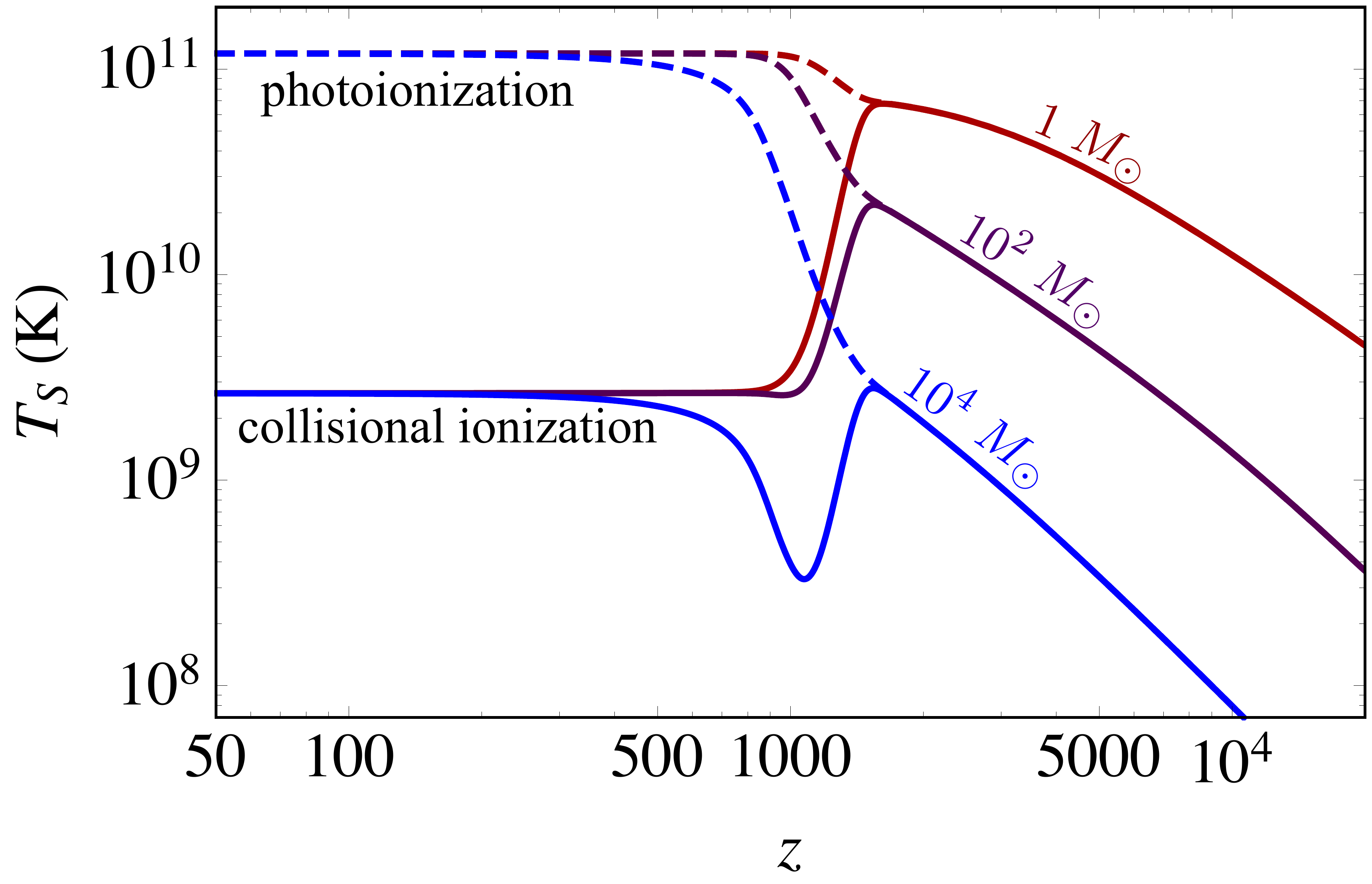}
\caption{Characteristic temperature of the accreting gas near the Schwarzschild radius, evaluated with the substitution $v_{\rm B} \rightarrow v_{\rm eff}$ as described in Section \ref{sec:velocities}. } \label{fig:Ts}
\end{figure}

Equations \eqref{eq:Y-final}, \eqref{eq:T-of-rho}, \revision{\eqref{eq:chi-general}} and \eqref{eq:tau_fit} constitute the second main result of this Section. They give the gas temperature near the BH horizon, accounting for Compton cooling and an arbitrary background ionization fraction, \revision{in the two limiting cases of collisional ionization or photoionization}. We show the temperature $T_{\rm S}$ as a function of redshift and PBH mass in Fig.~\ref{fig:Ts}. At high redshift, the temperature is suppressed by the strong Compton cooling. In the \revision{collisional ionization} case, once the Universe becomes neutral, some thermal energy is used in ionizing the gas, lowering $T_{\rm S}$ \change{by a factor up to $\sim 300$, corresponding to the radial extent of the ionization region.}

\subsection{Luminosity of an accreting black hole} \label{sec:luminosity}

The luminosity of the accreting BH arises mostly from Bremsstrahlung (free-free) radiation near the Schwarzschild radius. We show in Appendix \ref{sec:fb} that free-bound radiation is negligible with respect to free-free radiation. 

The frequency-integrated emissivity (in ergs/s/cm$^3$) of a fully-ionized thermal electron-proton plasma can be written in the general form (see e.g.~Ref.~\cite{Stepney_83})
\beq
j^{\rm ff} = n_e^2 ~\alpha c \sigma_{\rm T}  T ~ \mathcal{J}(T/m_e c^2), \label{eq:j_ff}
\eeq
where $\alpha$ is the fine-structure constant and $\mathcal{J}(X)$ is a dimensionless function. Ref.~\cite{Svensson_82} provide a simple fitting formula for the $e-p$ free-free emissivity, accurate to a few percent, and Ref.~\cite{Nozawa_09} provide a sub-percent accuracy code for the $e-e$ free-free emissivity. We fit the sum of the two within a few percent by the following analytic approximation, generalizing that of Ref.~\cite{Svensson_82}:
\barr
\mathcal{J}(X) \approx 
\begin{cases}
\frac4{\pi} \sqrt{2/\pi} X^{-1/2} \left(1 + 5.5 X^{1.25} \right),  & X < 1,\\[6pt]
\frac{27}{2 \pi} \left[\ln(2 X \rme^{- \gamma_{\rm E}} + 0.08) + \frac43\right],  & X > 1,~~~~~
\end{cases}
\earr
where $\gamma_{\rm E} \approx 0.577$ is Euler's Gamma constant. Assuming the plasma is optically thin (which we show explicitly in Appendix \ref{sec:opt_thin}), the luminosity is then obtained by integrating the emissivity over volume, $L = \int 4 \pi r^2 dr j$. Let us note that this purely Newtonian expression does not properly account for relativistic effects which become relevant near the horizon \cite{Shapiro_73}; they results in order-unity corrections which are below our theoretical uncertainty.

\begin{figure}
\includegraphics[width = \columnwidth]{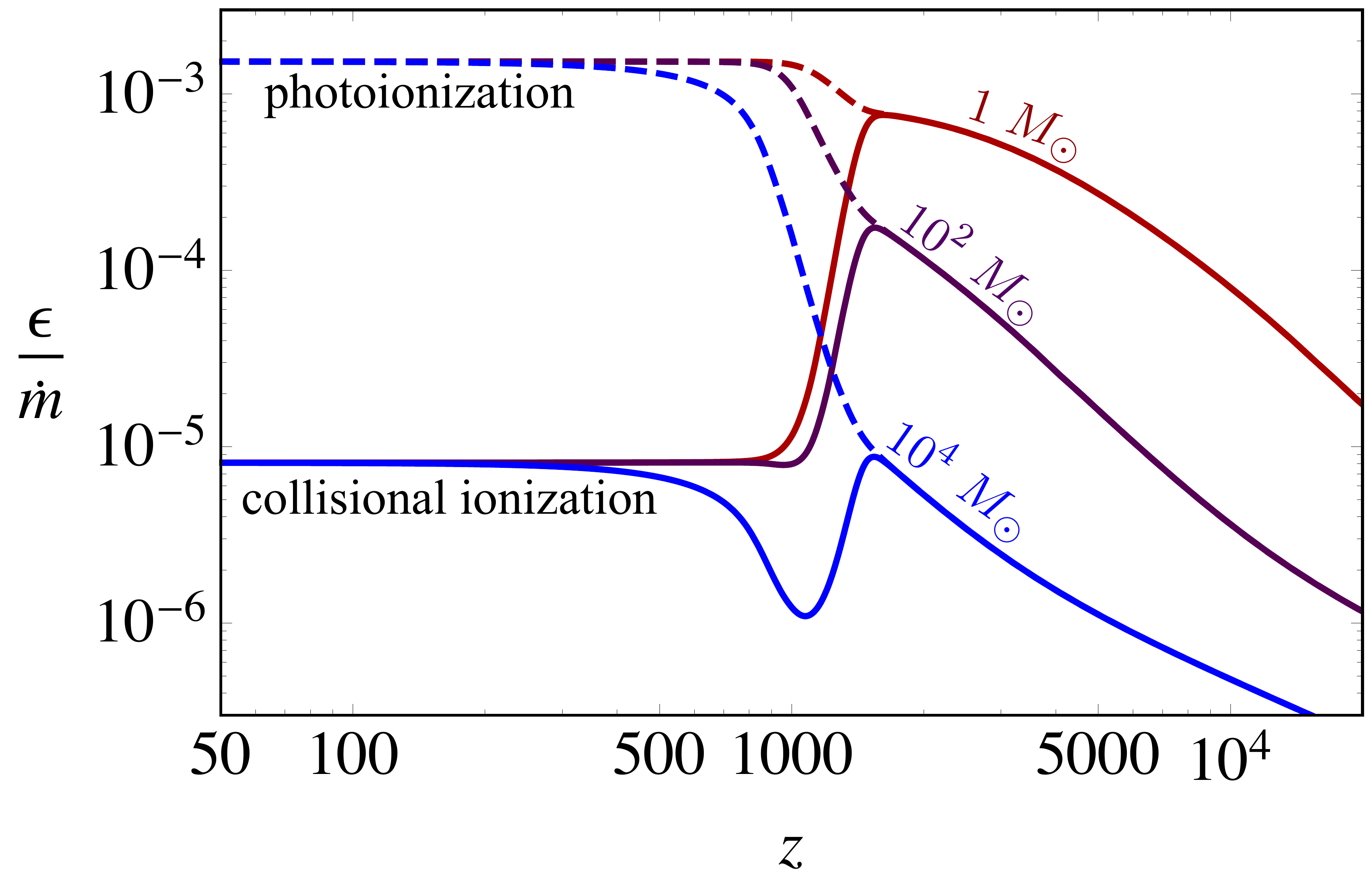}
\caption{\revision{Radiative efficiency $\epsilon \equiv L/\dot{M} c^2$ divided by the dimensionless accretion rate $\dot{m} \equiv \dot{M} c^2/L_{\rm Edd}$, evaluated with the substitution $v_{\rm B} \rightarrow v_{\rm eff}$ as described in Section \ref{sec:velocities}.}} \label{fig:epsilon}
\end{figure}

Near the Schwarzschild radius the gas is in free-fall, $|v| \approx c \sqrt{r_{\rm S}/r}$, and the electron density results from the mass-conservation equation:
\beq
n_e = \frac{\dot{M}}{4 \pi m_p r^2 |v|} = \frac{\dot{M}}{4 \pi m_p r_{\rm S}^2 c} (r/r_{\rm S})^{-3/2}. \label{eq:ne}
\eeq
The radial dependence of the temperature near the horizon depends on $T/m_e c^2$. For the range of temperature considered we find that $ 0.8 \lesssim - d \ln (T \mathcal{J})/d \ln r \lesssim 1.1$. Approximating $T \mathcal{J}(T) \propto r^{-1}$, \change{we therefore get}
\beq
L \approx \alpha \frac{T_{\rm S} }{m_p c^2}\mathcal{J}(T_{\rm S}) \frac{\dot{M} c^2}{L_{\rm Edd}} \dot{M} c^2, \label{eq:Luminosity}
\eeq
where we recall that the Eddington luminosity is 
\beq
L_{\rm Edd} \equiv \frac{4 \pi G M m_p c}{\sigma_{\rm T}}. \label{eq:Ledd}
\eeq
With this we see that the radiative efficiency $\epsilon \equiv L/\dot{M} c^2$ is proportional to $
\dot{m} \equiv \dot{M}c^2/L_{\rm Edd}$, with
\beq
\frac{\epsilon}{\dot{m}} \approx \alpha \frac{T_{\rm S} }{m_p c^2}\mathcal{J}(T_{\rm S}).
\eeq
The highest temperature, hence efficiency, is achieved when Compton cooling is negligible and the background is fully ionized, in which case \revision{we find $T_{\rm S} \approx 10^{11}$ K and $\epsilon/\dot{m} \approx 0.0015$. This is nearly} one order of magnitude below the value $\epsilon/\dot{m} = 0.011$ assumed in ROM, and is further suppressed at most times, as we show in Fig.~\ref{fig:epsilon}. 

\subsection{Accounting for BH velocities} \label{sec:velocities}

All of the calculations so far assume perfectly spherically-symmetric accretion. In practice, the accreting PBHs are moving with respect to the ambient gas with some velocity $v$.

It is not at all clear what the best way is to account for the black hole peculiar velocity without performing a full time-dependent hydrodynamical simulation. Bondi and Hoyle \cite{Bondi_44} studied analytically accretion on a point mass moving highly supersonically, and found $\dot{M} \approx 2.5 \pi (G M)^2 \rho_{\infty}/v^3$. Inspired by this result, Bondi \cite{Bondi_52} suggested substituting the sound speed at infinity $c_s$ by $\sqrt{c_s^2 + v^2}$ in the accretion rate, which, he argued, ought to give the correct order of magnitude for the result. Though this provides a prescription for the accretion rate, it is not clear how to self-consistently account for relative velocities in the estimate of the gas temperature. 
For definiteness, and lacking a better theory, we shall approximate the effect of relative velocities by substituting $v_{\rm B}^2 \rightarrow v_{\rm B}^2 + v^2$ throughout the calculation. This is equivalent to substituting $T_{\infty} \rightarrow T_{\infty} + m_p v^2/(1 + \overline{x}_e)$. The same route was followed in ROM.

The relative velocity $v$ is comprised of two pieces: a Gaussian linear contribution on large scales, $v_{\rm L}$, whose power spectrum and variance can be extracted from linear Boltzmann codes, and a small-scale contribution due to non-linear clustering of PBHs, $v_{\rm NL}$. We shall not consider the latter here, but point out that it would further suppress the effect of PBHs on the CMB.

If PBHs make up the dark matter, the linear velocity $v_{\rm L}$ is nothing but the relative velocity of baryons and dark matter. After kinematic decoupling at $z \approx 10^3$, dark matter and baryons fall in the same gravitational potentials on scales larger than the baryon Jeans scale and hence $v_{\rm L} \propto 1/a$, independent of scale \cite{Tseliakhovich_10}. Before then, however, the relative velocity has a more complex time and scale-dependence since baryons undergo acoustic oscillations while the dark-matter overdensities grow. Ref.~\cite{Dvorkin_14} explicitly compute $\langle v_{\rm L}^2 \rangle$ as a function of time and find that it is mostly constant for $z \gtrsim 10^3$ (see their Fig.~1). Since, as we shall see, most of the effect of accreting PBHs on the CMB takes place after decoupling, we need not have a very precise estimate of $v_{\rm L}$ before then, and assume the following simple redshift dependence:
\beq
\langle v_{\rm L}^2 \rangle^{1/2}  \approx  \min\left[1, z/10^3\right] \times 30 ~\textrm{km/s}. \label{eq:v_rms}
\eeq
\change{Let us point out that the relative velocity adopted in ROM is quite different from what we use here (see their Fig.~2); in particular they under-estimate it for $z \gtrsim 200$, leading to an over-estimate of the accretion rate.}

\begin{figure}
\includegraphics[width = \columnwidth]{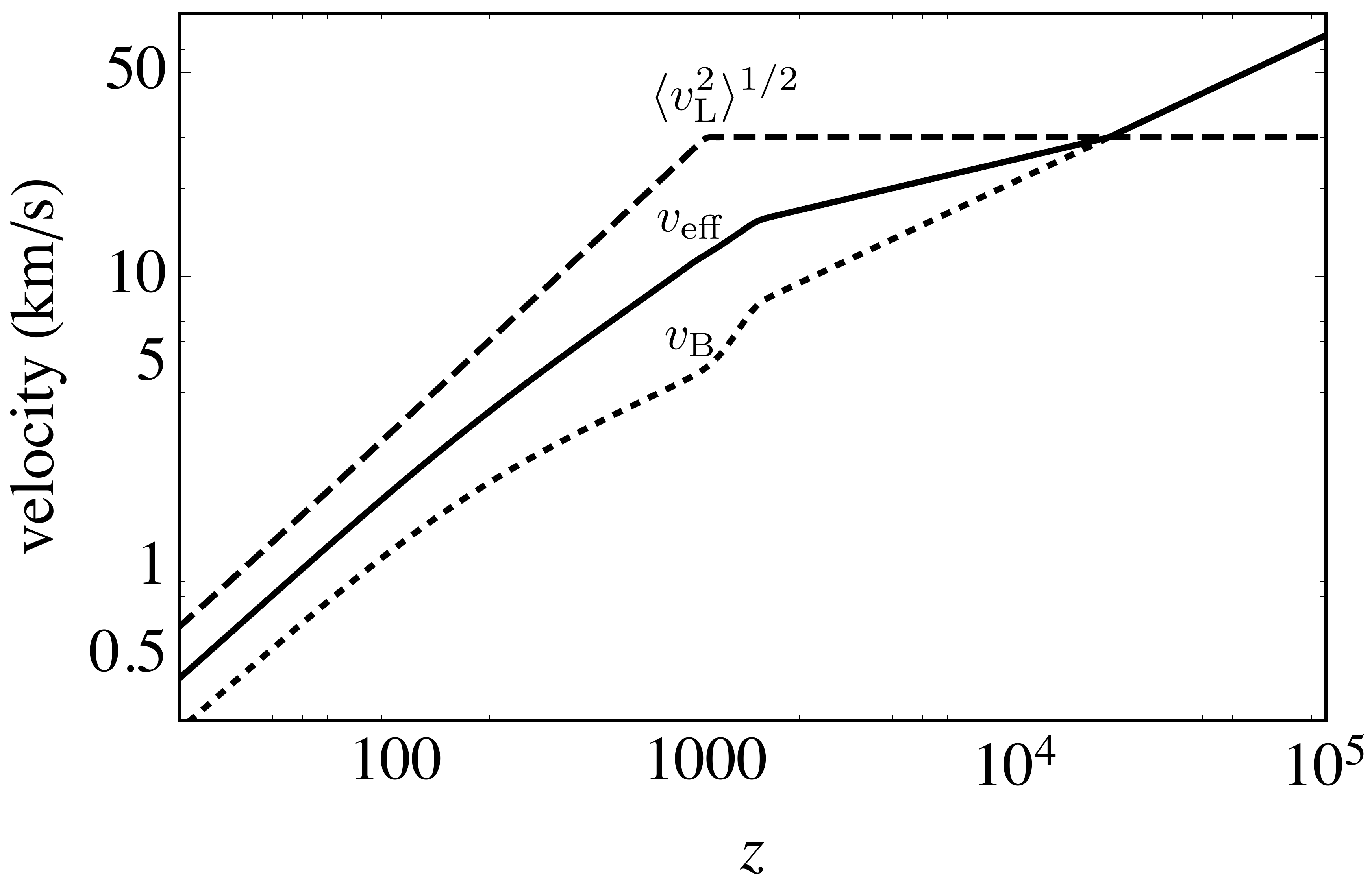}
\caption{Characteristic velocities in the problem at hand: the isothermal sound speed $v_{\rm B}$ (dotted), rms BH-baryon relative velocity $\langle v_{\rm L}^2\rangle^{1/2}$ (dashed) and effective velocity $v_{\rm eff}$ defined in Eq.~\eqref{eq:veff} (solid), used in Figures \ref{fig:beta_gamma}, \ref{fig:lambda}, \ref{fig:Ts} and \ref{fig:epsilon} to illustrate characteristic values of intermediate quantities.} \label{fig:velocities}
\end{figure}

\begin{figure}
\includegraphics[width = \columnwidth]{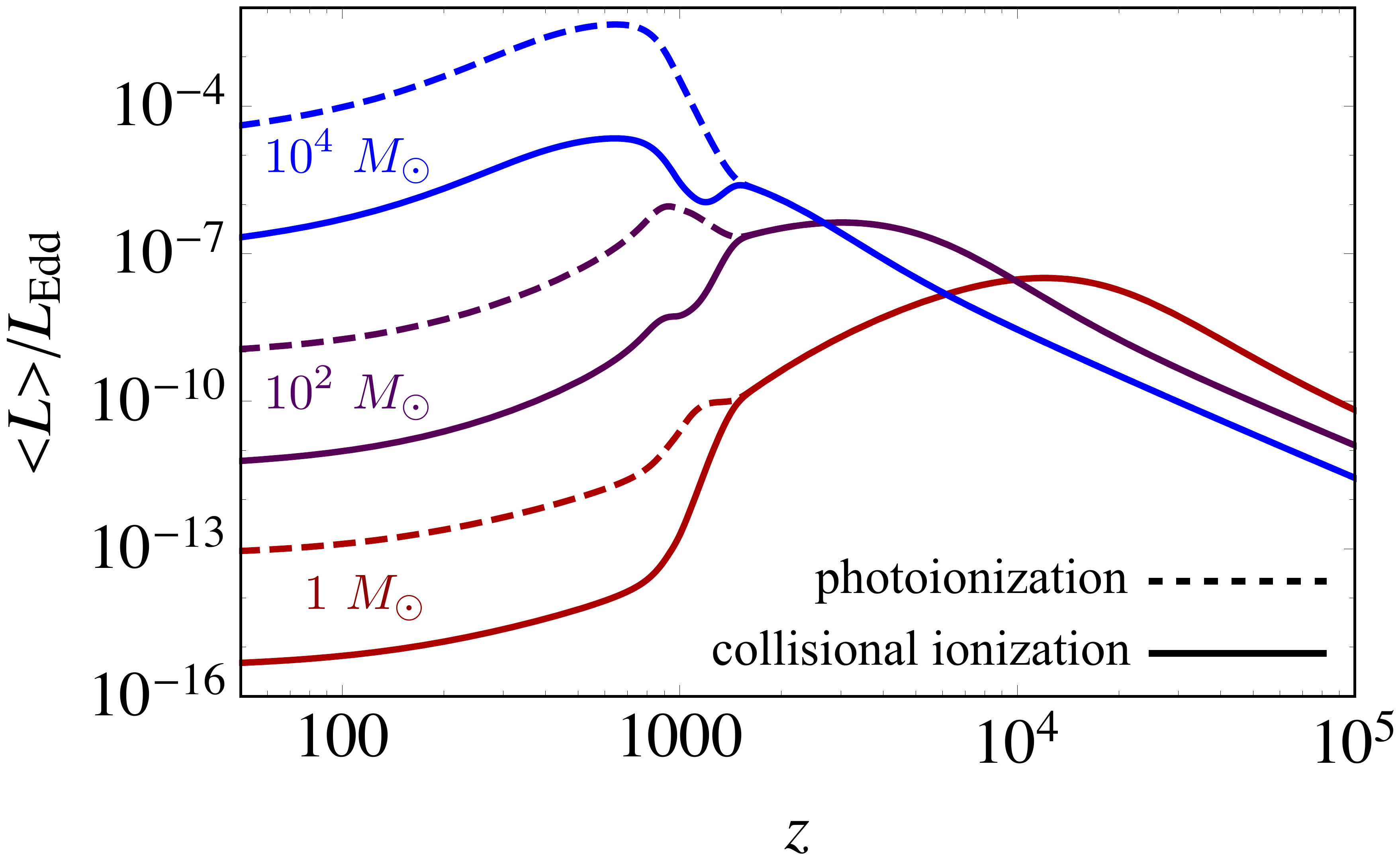}
\caption{Luminosity of accreting PBHs as a function of redshift, averaged over the Gaussian distribution of large-scale relative velocities.} \label{fig:luminosity}
\end{figure}

As we saw in Section \ref{sec:luminosity}, the BH luminosity is quadratic in the accretion rate, and therefore, in the standard Bondi case, proportional to $(v_{\rm B}^2 + v_{\rm L}^2)^{-3}$. The total energy injected in the plasma is obtained by averaging the BH luminosity over the Gaussian distribution of relative velocities. We define\footnote{This is equivalent to the quantity $\langle v_{\rm eff} \rangle_A$ in ROM.} $v_{\rm eff} \equiv \langle ( v_{\rm B}^2 + v_{\rm L}^2)^{-3} \rangle^{-1/6}$. It has the following approximate limits:
\beq
v_{\rm eff} \approx \begin{cases} \sqrt{v_{\rm B} \langle v_{\rm L}^2 \rangle^{1/2}}, \ \ &v_{\rm B} \ll \langle v_{\rm L}^2 \rangle^{1/2}\\
v_{\rm B}, \ \ &v_{\rm B} \gg \langle v_{\rm L}^2 \rangle^{1/2}
\end{cases} \label{eq:veff}
\eeq
We show $v_{\rm B}$, $\langle v_{\rm L}^2 \rangle^{1/2}$ and $v_{\rm eff}$ in Fig.~\ref{fig:velocities}. Figures \ref{fig:beta_gamma}, \ref{fig:lambda}, \ref{fig:Ts} and \ref{fig:epsilon} where all obtained by setting $v_{\rm B} \rightarrow v_{\rm eff}$, in order to illustrate the characteristic accretion rate and radiative efficiency. The final result of this Section is the luminosity of accreting PBHs, \emph{averaged over the distribution of relative velocities}, which we show in Fig.~\ref{fig:luminosity}. We emphasize that to obtain $\langle L \rangle$, we have replaced $v_{\rm B} \rightarrow \sqrt{v_{\rm B}^2 + v_{\rm L}^2}$ throughought the calculation, and then averaged the luminosity over the \change{three-dimensional Gaussian distribution of $\bs{v}_{\rm L}$.} 

\section{Local radiation feedback} \label{sec:feedback}

\revision{Before estimating the effect of the PBH radiation on the global thermal and ionization history, let us first examine whether it can affect the local accretion flow itself.}

\subsection{Local thermal feedback} \label{app:feedback}

\begin{figure}
\includegraphics[width = \columnwidth]{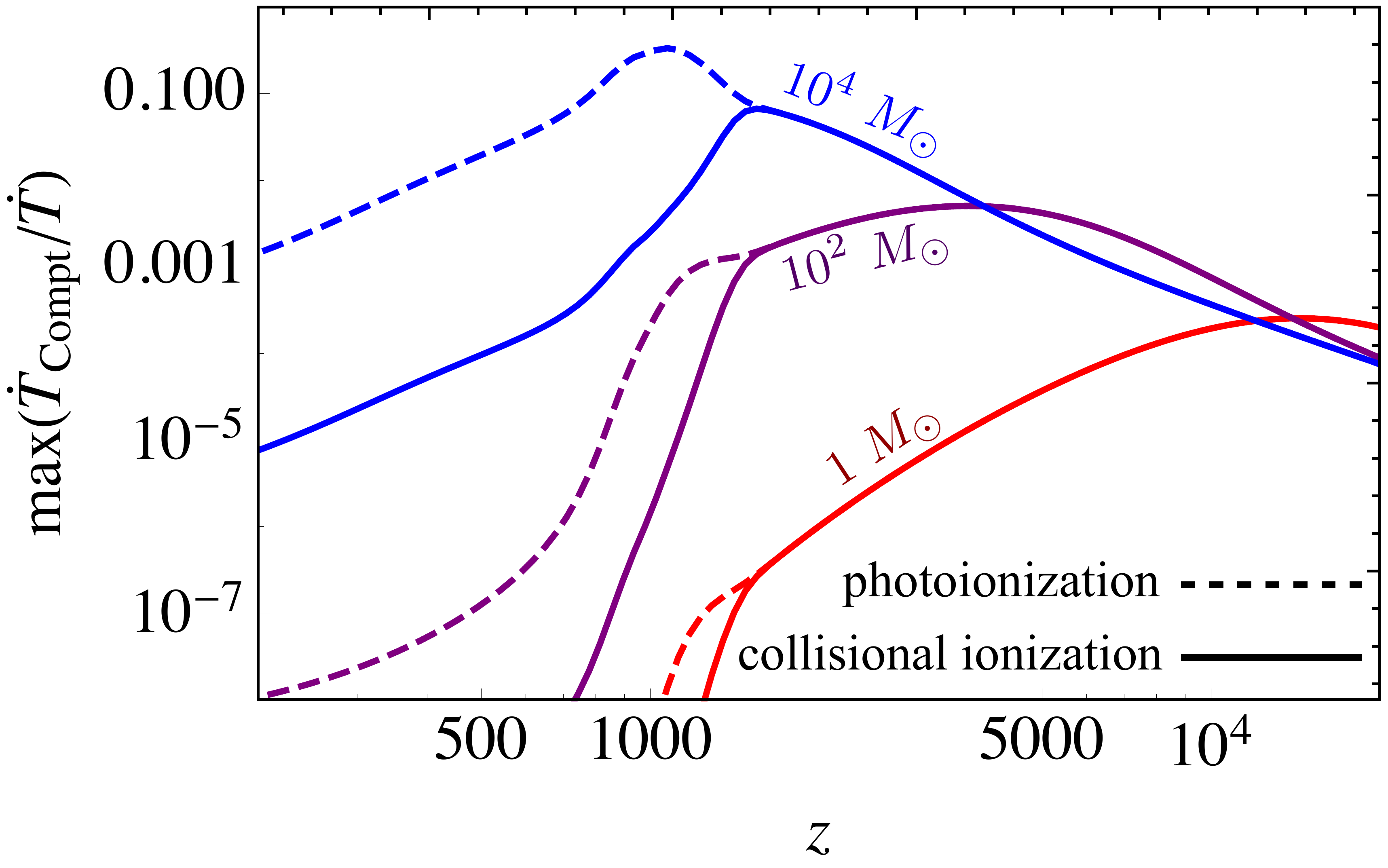}
\caption{Estimated maximum fractional importance of local thermal feedback from Compton heating by the PBH radiation.} \label{fig:feedback}
\end{figure}

Throughout the calculation we have neglected local Compton heating by the radiation produced by the accreting PBH. Here we discuss the validity of this assumption. The rate of energy injection per electron by Compton scattering with the PBH radiation is
\barr
\int dE \frac1{4 \pi r^2} \frac1{E} \frac{d L}{d E} \langle \sigma \Delta E \rangle \approx 0.1~ \frac{\sigma_{\rm T} L}{4 \pi r^2},
\earr
where we used the approximation \eqref{eq:Edot-approx} for $\langle \sigma \Delta E\rangle$. Hence the rate of Compton heating by the PBH radiation is 
\beq
\dot{T}_{\textrm{Compt}, L} \approx \frac{2}{3} \frac{x_e}{1 + x_e} 0.1~ \frac{\sigma_{\rm T} L}{4 \pi r^2}. 
\eeq
We need to compare this rate to the largest of the Compton cooling rate by CMB photons and the rate of adiabatic heating:
\barr
\dot{T}_{\rm Compt, cmb}  &\equiv& \frac{8}{3} \frac{x_e}{1 + x_e} \sigma_{\rm T} \frac{\rho_{\rm cmb} T_{\rm cmb}}{m_e c},\\
\dot{T}_{\rm ad} &\approx& T \frac{|v|}{r}.
\earr
If $\gamma\gg 1$ the latter two rates are approximately equal at $r_* \approx \gamma^{-2/3} r_{\rm B}$, adiabatic heating being dominant for $r \lesssim r_*$ and Compton cooling by CMB photons for $r \gtrsim r_*$ (see Section \ref{sec:compt_cool}). For $r < r_*$, $T \propto 1/r$ and $|v| \propto 1/r^{1/2}$ so $\dot{T}_{\textrm{Compt}, L}/\dot{T}_{\rm ad} \propto r^{1/2}$. For $r > r_*$, $\dot{T}_{\textrm{Compt}, L}/\dot{T}_{\rm Compt, cmb} \propto r^{-2}$. Therefore the impact of thermal feedback is maximized at $r \approx r_*$. If $\gamma \ll 1$, then we only need to compare the Compton heating rate to adiabatic cooling, at the Bondi radius where this ratio is maximized. We see that for arbitrary $\gamma$ the relevant radius at which to compare Compton heating to adiabatic cooling is $r \approx r_{\rm B}/(1+ \gamma^{2/3})$, where $T \approx T_{\rm cmb}$ in both cases. After some algebra we arrive at 
\beq
\max\left[\frac{\dot{T}_{\textrm{Compt}, L}}{\dot{T}}\right] \approx 0.07  \frac{x_e}{1 + x_e}  \frac{L}{L_{\rm Edd}} \frac{v_{\rm B}}{c} \frac{m_p c^2}{T_{\rm cmb}} \sqrt{1 + \gamma^{2/3}}.
\eeq
We show this ratio in Fig.~\ref{fig:feedback}, where we see that it is always less than unity for $M \leq 10^4 M_{\odot}$. We can therefore safely neglect local thermal feedback for the mass range we consider.

\revision{\subsection{Local ionization feedback}} \label{sec:ion_feedback}

\revision{Througout the paper we have computed all relevant quantities in both the ``collisional ionization" and the ``photoionization" limits. In the former case, we assumed that the radiation field from the accreting BH does not affect the ionization state of the gas in the immediate vicinity of the BH, so the gas gets eventually gets collisionally ionized, which reduces its temperature near the horizon. In the latter case, we assumed that the neighboring gas is fully photoionized. We now show that neither case is accurate and that within the adopted model, the level of feedback is expected to be somewhat intermediate between the two. To do so, we estimate the extent of the photoionized region (the Str\"omgren sphere) around an accreting PBH, in the absence of collisional ionizations. Following the standard derivation (see e.g. Ref.~\cite{Osterbrock_06}), 
\beq
\int_0^{\infty} 4 \pi r^2 dr n_e n_p \alpha_{\rm B}(T) =  \int_{\nu_0}^{\infty} d \nu \frac{L_{\nu}}{h \nu}, 
\eeq
where $\nu_0 = 13.6 \, \textrm{eV}/h$ is the ionization threshold and $\alpha_{\rm B}$ is the case-B recombination coefficient. This equation states that the total rate of recombinations is equal to the emission rate of ionizing photons. Note that it does not depend on the exact shape of the photoionization cross section (and in particular also accounts for ionizations by inelastic Compton scattering at high energies).}

\revision{Now we assume that the gas is fully ionized up to a radius $R$, after which it quickly becomes neutral. We also approximate $\alpha_{\rm B}(T) \propto T^{-q}$. Finally, in the free-fall limit, $n_e \propto 1/r^{3/2}$ and $T \propto 1/r$, implying
\barr
\alpha_{\rm B}(T) = \alpha_{\rm B, \rm ion} (r/r_{\rm ion})^q,  
\earr
where $\alpha_{\rm B, \rm ion} \equiv \alpha_{\rm B}(T_{\rm ion})$ and $r_{\rm ion}$ is the radius at which $T = T_{\rm ion} \equiv 1.5 \times 10^4$ K. Using Eq.~\eqref{eq:ne} we arrive at
\beq
\int_0^{\infty} 4 \pi r^2 dr n_e n_p \alpha_{\rm B}(T) = \frac{\dot{M}^2 \, \alpha_{\rm B, \rm ion}}{4 \pi  (m_p c)^2 r_{\rm S}}  \frac{(R/r_{\rm ion})^q}{q}.
\eeq
To compute the number of ionizing photons, we assume an approximately flat spectrum $L_{\nu} \approx L/\nu_{\max}$ for $\nu \leq \nu_{\max} \equiv T_{\rm S}/h$, so that
\beq
\int_{\nu_0}^{\infty} d \nu \frac{L_{\nu}}{h \nu} \approx \frac{L}{T_{\rm S}} \ln(T_{\rm S}/h \nu_0).
\eeq
Using Eq.~\eqref{eq:Luminosity} for $L$, we arrive at the following expression for the radius $R$, that does not explicitly depend on the luminosity or the accretion rate, but does depend weakly on $T_{\rm S}$, the temperature at the horizon:
\beq
\frac{R}{r_{\rm ion}} \approx \left[\frac{\alpha c \sigma_{\rm T}}{\alpha_{\rm B, \rm ion}}\mathcal{J}(T_{\rm S})  \ln(T_{\rm S}/h\nu_0) \right]^{1/q}.
\eeq
From Ref.~\cite{Pequignot_91} we get $\alpha_{\rm B, \rm ion} \approx 1.8  \times 10^{-13}$ cm$^{-3}$ s$^{-1}$, with a local power law $q \approx 0.86$, hence
\beq
\frac{R}{r_{\rm ion}}  \approx 2 \times 10^{-4} \left[\mathcal{J}(T_{\rm S}) \ln(T_{\rm S}/h\nu_0) \right]^{1.16}.
\eeq
Let us now consider the two limiting regimes once the background ionization fraction drops significantly below unity. Assuming the gas is photoionized by the radiation field rather than collisionally ionized, we found $T_{\rm S} \approx 10^{11}$ K at low redshift, implying
\beq
\frac{R}{r_{\rm ion}} \approx 0.1  \ \ \ (T_{\rm S} = 10^{11} \, \rm K).
\eeq
Since this is less than unity, this implies that assuming ionizations proceed exclusively through photoionizations is not self-consistent, as the radiation from the BH cannot photoionize the gas all the way to $r_{\rm ion}$. Let us notice that this implies a fortiori that the photoionized region does not extend to the outermost region where $T \approx T_{\rm cmb}$, and that we are hence justified in assuming  $x_e = \overline{x}_e$ there.}

\revision{If we instead take the ``collisional ionization" limit, for which $T_{\rm S} \approx 3 \times 10^9$ K, at low redshift, we get
\beq
\frac{R}{r_{\rm ion}} \approx 0.02  \ \ \ (T_{\rm S} = 3 \times 10^{9} \, \rm K).
\eeq
This radius is larger than the innermost edge of the collisional ionization region, which we found to be $\sim 0.003 \, r_{\rm ion}$. This implies that it is also not self-consistent to assume that the gas is exclusively collisionally ionized, as the photoionization region from the resulting radiation field would extend inside the collisional ionization region.}

\revision{We therefore conclude that neither approximation is self-consistent, and that the actual luminosity (within our assumed spherical accretion model) is intermediate between these two limiting cases. We now move on to compute the global effects of the the PBH luminosity on the background gas.}

\section{Energy deposition in the plasma} \label{sec:deposition}

\subsection{Total energy deposition rate}

Assuming PBHs make a fraction $f_{\rm pbh}$ of the dark matter, the volumetric rate of energy \emph{injection} (in ergs/cm$^3$/s) by accreting PBHs is 
\beq
\dot{\rho}_{\rm inj} = f_{\rm pbh} \frac{\rho_{\rm dm}}{M} \langle L\rangle.
\eeq
This energy is injected in the form of a nearly flat photon spectrum (i.e.~the free-free luminosity per frequency interval $dL/d \nu$ is approximately constant), up to maximum energy $E_{\max} \approx T_{\rm S}$, typically $\sim 0.2$ MeV for $z \lesssim 10^3$ and up to $\sim 6$ MeV at higher redshifts. 

What is relevant for cosmological observables is the volumetric rate of energy \emph{deposited} in the plasma (in the form of heat or ionizations), which we denote by $\dot{\rho}_{\rm dep}$. The two rates are not necessarily equal, unless energy is deposited on-the-spot. 

At the characteristic energies considered, the dominant photon cooling process is inelastic Compton scattering off electrons, whether bound or free \cite{Chen_04, Slatyer_09}. In principle, in order to obtain the energy deposition rate one should solve for the time evolution of the photon distribution, as well as that of the secondary high-energy electrons resulting from Compton scattering. To simplify matters we shall assume that the latter deposit their energy on-the-spot, so we only need to follow the photon distribution $\mathcal{N}_E$ (in photons/cm$^3$/erg). 

\begin{figure}
\includegraphics[width = \columnwidth]{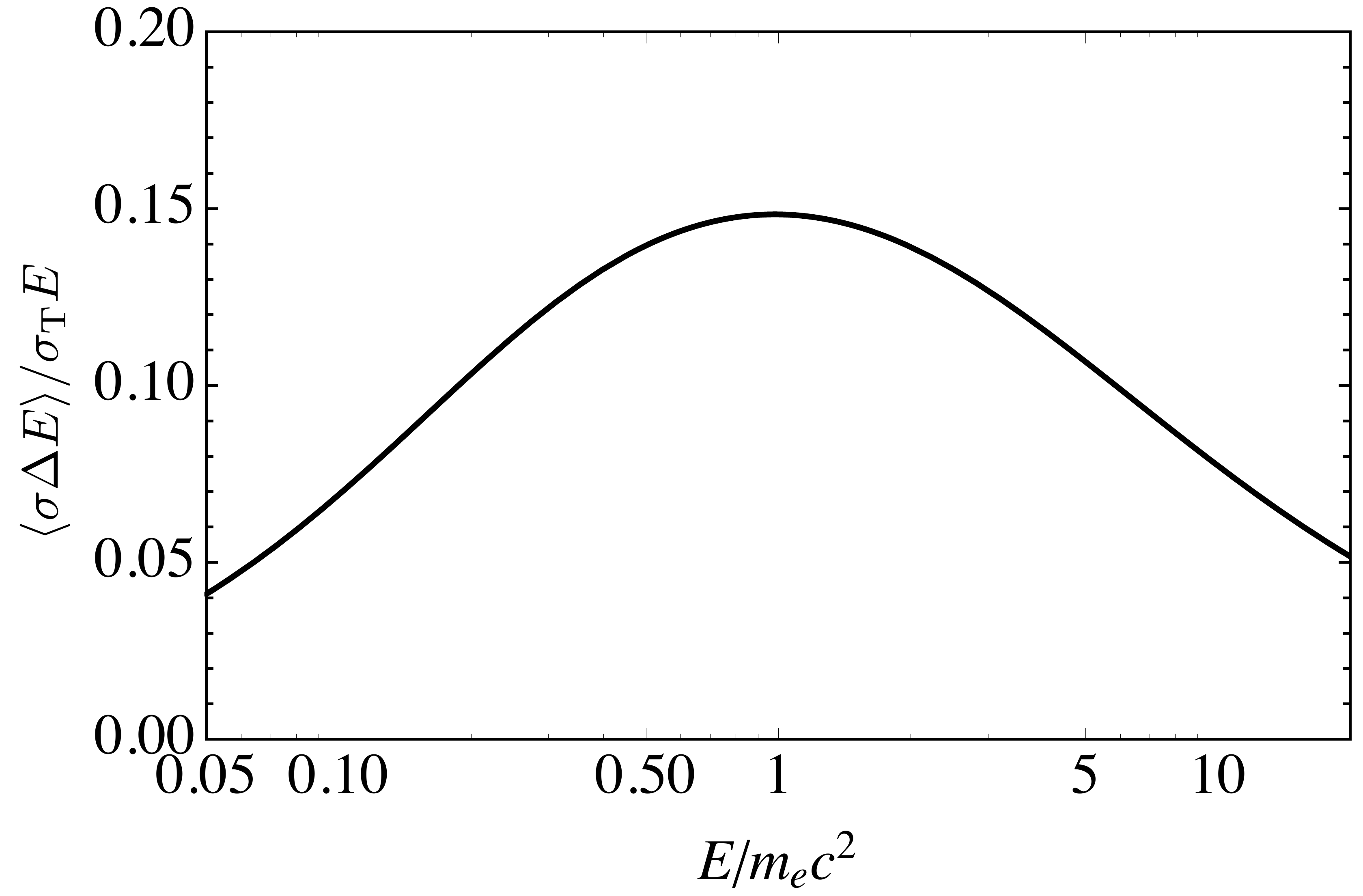}
\caption{Ratio of the cross-section-averaged energy loss per Compton scattering event to $\sigma_{\rm T} E$, as a function of photon energy.} \label{fig:sigmaE}
\end{figure}

The differential scattering cross section for Compton scattering is \cite{Slatyer_09}
\barr
\frac{d \sigma(E)}{d E'} &=& \frac38 \sigma_{\rm T} \frac{m_e c^2}{E^2}\nonumber\\
&\times& \left[\frac{E'}{E} + \frac{E}{E'} - 1 + \left(1 + \frac{m_e c^2}{E} - \frac{m_e c^2}{E'}\right)^2 \right],  ~~~~\label{eq:cross-section}
\earr
where $E$ is the initial energy of the photon and $E'$ is its final energy, restricted to the range
\beq
E'_{\min}(E) \equiv \frac{E}{1 + 2 E/m_e c^2} \leq E' \leq E. \label{eq:E'min}
\eeq 
In principle the photon distribution $\mathcal{N}_E$ should be obtained by solving an integro-differential Boltzmann equation. To simplify, we approximate the Boltzmann equation by the continuity equation\footnote{One can also think of this equation as a Fokker-Planck equation without a diffusion term.} 
\beq
a^{-2}\frac{d}{d t}(a^2 \mathcal{N}_E) \approx \frac1{E} \frac{d \dot{\rho}_{\rm inj}}{dE} +\frac{\partial}{\partial E}\left( \dot{\mathcal{E}}(E) \mathcal{N}_E \right), \label{eq:continuity}
\eeq
where $d/dt \equiv \partial/\partial t - H E \partial/\partial E$ is the derivative along the photon geodesics and
\beq
\dot{\mathcal{E}}(E) \equiv \overline{n}_{\rm H} c \langle \sigma \Delta E \rangle
\eeq
is the rate of energy loss due to Compton scattering, where
\beq
\langle \sigma \Delta E \rangle \equiv \int_{E'_{\min}(E)}^E d E' \frac{d \sigma(E)}{dE'}(E - E').
\eeq
The $a^2$ factors in Eq.~\eqref{eq:continuity} ensure that $\mathcal{N}_E \propto a^{-2}$ in the absence of the source and collision terms, and the form of the differential operator for Compton scattering explicitly conserves the number of photons. We show the ratio $\langle \sigma \Delta E \rangle / \sigma_{\rm T} E$ in Fig.~\ref{fig:sigmaE}. We see that for the range of energies considered, within a factor of 2 at most, 
\beq
\langle \sigma \Delta E \rangle \approx 0.1 ~ \sigma_{\rm T} E. \label{eq:Edot-approx}
\eeq
The factor of 0.1 can be understod as follows. For $E \gtrsim m_e c^2$, photons lose most of their energy in each scattering event, but the Compton cross-section is suppressed with respect to the Thomson limit. For $E \lesssim m_e c^2$, the Compton cross section tends to the Thomson limit, but photons only lose a small fraction of their energy in each scattering event. 

Within our set of approximations, the differential energy deposition rate is
\beq
\frac{d \dot{\rho}_{\rm dep}}{d E} \approx \dot{\mathcal{E}}(E) \mathcal{N}_E \approx 0.1 ~\overline{n}_{\rm H} c \sigma_{\rm T} E \mathcal{N}_E.
\eeq
From Eq.~\eqref{eq:continuity} we find that this quantity satisfies the following equation
\barr
a^{-6}\frac{d}{dt}\left(a^6 \frac{d \dot{\rho}_{\rm dep}}{dE} \right) &\approx& 0.1~ \overline{n}_{\rm H} c \sigma_{\rm T} \nonumber\\
&\times&\left[ \frac{d \dot{\rho}_{\rm inj}}{dE} + E \frac{\partial}{\partial E}\left(\frac{d \dot{\rho}_{\rm dep}}{dE} \right)\right].~~~
\earr
Integrating over energies (and recalling that $d/dt = \partial/\partial t - H E \partial/\partial E$), we arrive at the following very simple differential equation for the total energy deposition rate:
\beq
a^{-7} \frac{d}{dt}(a^7 \dot{\rho}_{\rm dep}) \approx 0.1~ \overline{n}_{\rm H} c \sigma_{\rm T} (\dot{\rho}_{\rm inj} - \dot{\rho}_{\rm dep}). \label{eq:rho_dep}
\eeq
\revision{We compare and contrast our results to existing analytic calculations in Appendix \ref{app:comparison}.}

\revision{Physically, Eq.~\eqref{eq:rho_dep} implies that $\dot{\rho}_{\rm dep} \approx \dot{\rho}_{\rm inj}$ (i.e. that the energy is deposited ``on the spot") as long as the Compton cooling timescale $(0.1 c \sigma_{\rm T} \overline{n}_{\rm H})^{-1}$ is much shorter than the characteristic timescale over which $\dot{\rho}_{\rm inj}$ changes. Once this is no longer the case, the deposited energy rapidly decays as $1/a^7$. The Compton cooling timescale becomes longer than the Hubble timescale at $z \approx 200$. However $\dot{\rho}_{\rm inj}$ can change on a timescale significantly shorter than a Hubble time, in particular around recombination (see Fig.~\ref{fig:luminosity}), so $\dot{\rho}_{\rm dep}$ may deviate from $\dot{\rho}_{\rm inj}$ even earlier on.} 

\begin{figure}
\includegraphics[width = \columnwidth]{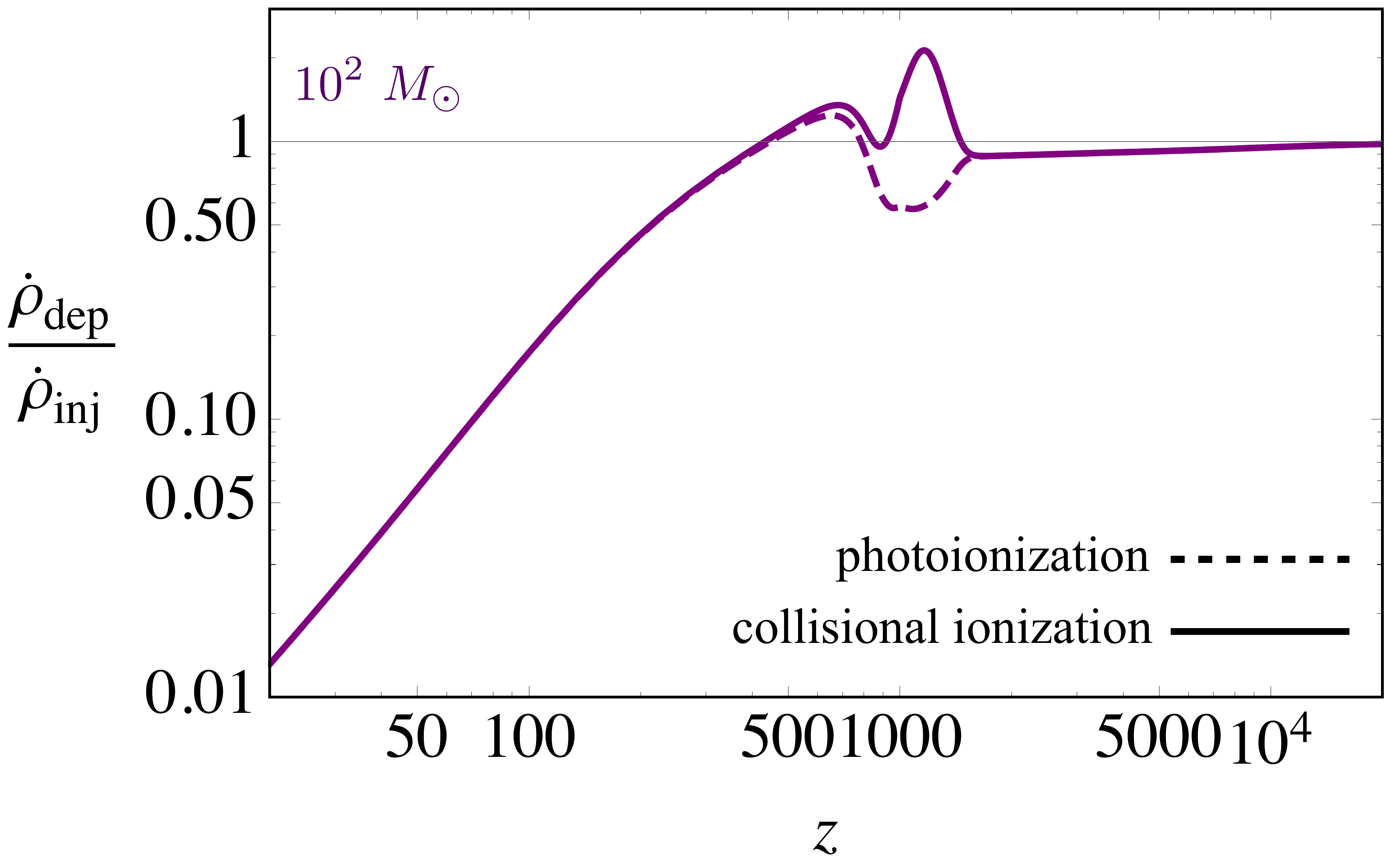}
\caption{Ratio of the energy deposition rate to the instantaneous energy injection rate (equivalent of the dimensionless efficiency $f(z)$ usually computed in the context of dark-matter annihilation), as a function of redshift. We only show the case $M = 10^2\, M_{\odot}$ as other cases are very similar.} \label{fig:f(z)}
\end{figure}

We show the ratio $\dot{\rho}_{\rm dep}/\dot{\rho}_{\rm inj}$ as a function of redshift \change{for a $10^2 \, M_{\odot}$ PBH} in Fig.~\ref{fig:f(z)}. Note that this is conceptually equivalent to the dimensionless efficiency $f(z)$ usually computed in the context of dark-matter annihilation (see e.g.~Ref.~\cite{Slatyer_09}). We see that this ratio \change{goes to unity at $z \gtrsim 10^3$}, and is suppressed for $z \lesssim 300$, as expected. Interestingly, in the \revision{collisional ionization} case, this ratio can actually be larger than unity around $z \sim 10^3$. This is due to the sharp decrease of the PBH average luminosity at recombination for $M \lesssim 10^2\, M_{\odot}$ (see Fig.~\ref{fig:luminosity}), hence of the instantaneous injected energy, and the non-negligible time-delay between injection and deposition already present at that redshift.

\subsection{Effect on the thermal and ionization histories}

To conclude this Section, we must describe how exactly the energy is deposited in the plasma. We follow the simple prescription of Ref.~\cite{Chen_04}, assuming that for a neutral gas the deposited energy is equally split among heating, ionizations and excitations, and rescale these fractions for arbitrary ionization fractions. We only consider the effect on hydrogen recombination for simplicity. Specifically, we take the following prescriptions for the additional rates of change of gas temperature, direct ionizations and excitations: 
\barr
\Delta \dot{T}_{\rm gas} &=& \frac2{3 n_{\rm tot}} \frac{1 + 2 x_e}{3} \dot{\rho}_{\rm dep}, \\
\Delta \dot{x}_e^{\rm direct} &=& \frac{1- x_e}{3} \frac{\dot{\rho}_{\rm dep}}{E_{\rm I} n_{\rm H}},\\
\Delta \dot{x}_2 &=&  \frac{1- x_e}{3} \frac{\dot{\rho}_{\rm dep}}{E_{\rm 2} n_{\rm H}},
\earr
where $n_{\rm tot}$ is the total number density of free particles, $x_2$ is the fraction of excited hydrogen and $E_2 \equiv 10.2$ eV is the first excitation energy (we assume that all excitations are to the first excited state for simplicity). Note that in our previous notation $x_e \equiv \overline{x}_e$ is the background ionization fraction and similarly $T_{\rm gas} \equiv T_{\infty}$.

\begin{figure}
\includegraphics[width = \columnwidth]{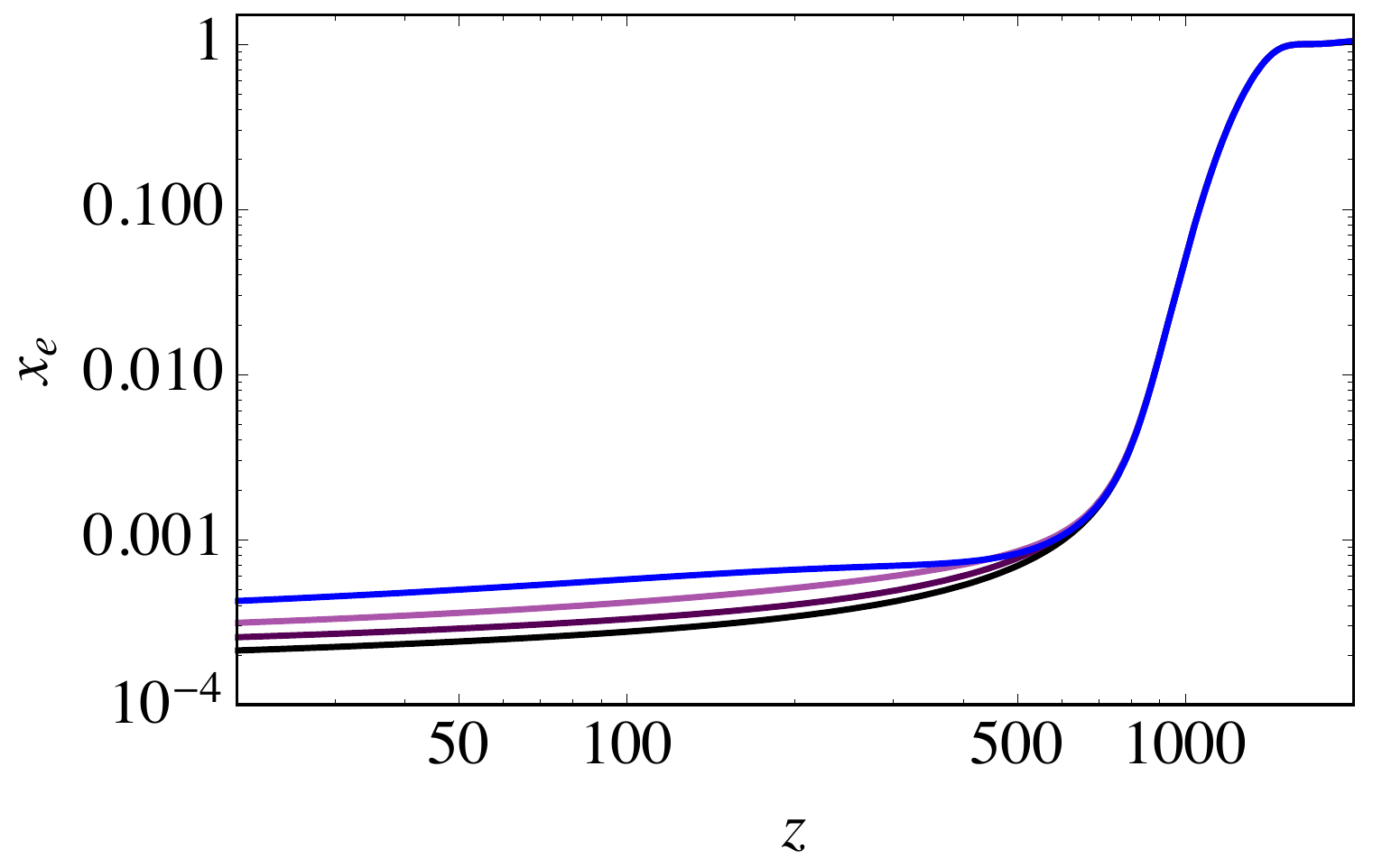}
\includegraphics[width = \columnwidth]{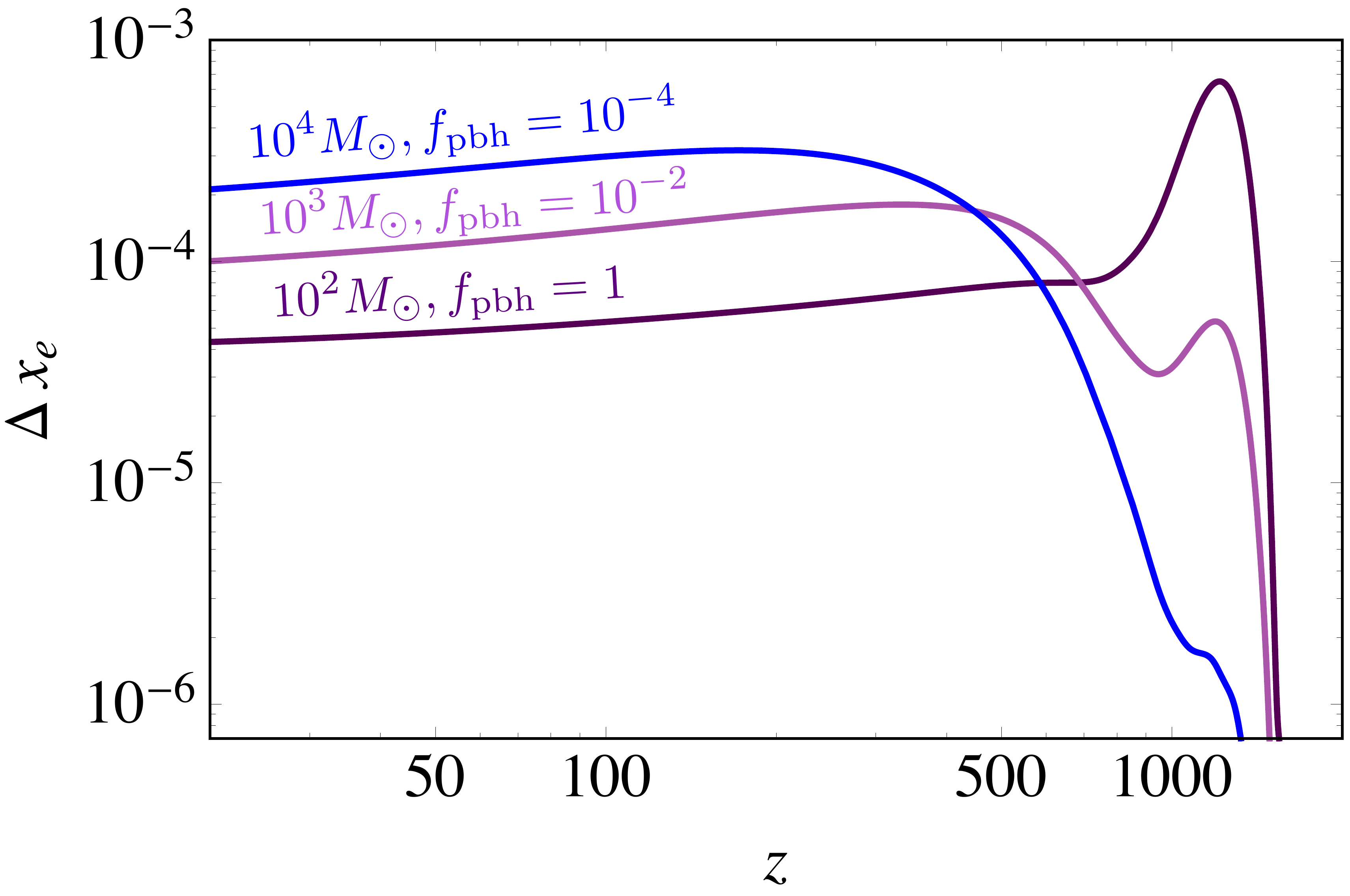}
\caption{\emph{Upper panel}: global free electron fraction $x_e(z)$ in the standard scenario (lower black curve), and accounting for PBHs with parameters $(M_{\rm pbh}/M_{\odot}, f_{\rm pbh}) = (10^2, 1), (10^3, 10^{-2}), (10^4, 10^{-4})$, in that order from bottom to top at low redshift. \emph{Lower panel}: change in the ionization history due to accreting PBHs \change{for the same parameters}. We only show the \revision{collisional ionization} case here.} \label{fig:xe}
\end{figure}

We implement these modifications in the recombination code \textsc{hyrec} \cite{YAH_10, YAH_11}. We self-consistently account for the heating of the gas into the PBH luminosity, i.e. account for the global feedback of PBHs. We show the resulting changes in the ionization history in Fig.~\ref{fig:xe}. Comparing with Fig.~3 of ROM, we see that we obtain a significantly smaller effect on the ionization history.

\section{Effect on the cosmic microwave background} \label{sec:cmb}

\subsection{CMB spectral distortions}

\subsubsection{Effect of global heating}

Energy deposited in the photon-baryon plasma at redshift $z \lesssim 2\times 10^6$ does not get fully thermalized, and results in distortions to the CMB spectrum. Depending on when the energy is deposited, the distortion generated is either a chemical potential ($\mu$-type) or a Compton-$y$ distortion. Their amplitudes are approximately given by (see e.g.~\cite{Chluba_16})
\barr
\mu &\approx& 1.4 \int_{5\times 10^4}^{2 \times 10^6} d \ln(1 + z) \frac{\dot{\rho}_{\rm dep}^{\rm heat}}{H \rho_{\rm cmb}},\\
y &\approx& \frac14 \int_{200}^{5 \times 10^4} d \ln(1 + z) \frac{\dot{\rho}_{\rm dep}^{\rm heat}}{H \rho_{\rm cmb}}. \label{eq:y}
\earr
The relevant ratio is therefore that of the volumetric rate of heat deposition per Hubble time to the CMB photon energy density:
\barr
\frac{\dot{\rho}_{\rm dep}^{\rm heat}}{H \rho_{\rm cmb}} &\approx& \frac{1 + 2 x_e}{3}  \frac{\dot{\rho}_{\rm dep}}{\dot{\rho}_{\rm inj}} f_{\rm pbh}  \frac{\rho_{\rm dm}}{\rho_{\rm cmb}} \frac{\langle L \rangle}{H M}\nonumber\\
&\approx& 4\times 10^{-4}~ \frac{1 + 2 x_e}{3} \frac{\dot{\rho}_{\rm dep}}{\dot{\rho}_{\rm inj}} f_{\rm pbh} \frac{\langle L\rangle }{L_{\rm Edd}} \frac{(z_{\rm eq}/z)^3}{\sqrt{1 + z_{\rm eq}/z}},~~~~~
\earr
where $z_{\rm eq} \approx 3400$ is the redshift of matter-radiation equality. In the $\mu$-era $z \gtrsim 5 \times 10^4$, we have $x_e \rightarrow 1$ (neglecting Helium), $\dot{\rho}_{\rm dep} = \dot{\rho}_{\rm inj}$, and $z \gg z_{\rm eq}$, and we arrive at
\beq
\mu \leq 6 \times 10^{-8} f_{\rm pbh} ~ \underset{z \geq 5 \times 10^4}{\max}\left(\frac{\langle L\rangle}{L_{\rm Edd}}\right). 
\eeq
This is always significantly below the sensitivity of FIRAS \cite{Fixen_96}, and would be within the reach of proposed spectral distortion experiments such as PIXIE \cite{Kogut_11} only if PBHs radiated near the Eddington luminosity. In practice, $L \ll L_{\rm Edd}$ at all times (see Fig.~\ref{fig:luminosity}), hence we conclude that accreting PBHs are not and will never be detectable through $\mu$-type spectral distortions.

The $y$-parameter integral \eqref{eq:y} is dominated by the lower redshift cutoff $z \approx 200$ corresponding to the thermal decoupling of gas and CMB photons. Since the luminosity is a slowly varying function near $z \approx 200$ and  $\dot{\rho}_{\rm dep} \approx \dot{\rho}_{\rm inj}$, we find
\beq
y \approx 0.02 ~ f_{\rm pbh} \frac{\langle L\rangle}{L_{\rm Edd}}\Big{|}_{z \approx 200}.
\eeq
From Fig.~\ref{fig:luminosity}, we see that for the mass range considered $M \leq 10^4\, M_{\odot}$ this is always below the sensitivity of FIRAS \cite{Fixen_96}. For $M = 10^4\, M_{\odot}$, the $y$-parameter may be as large as $y \sim 2 \times 10^{-7} f_{\rm pbh}$. This is within the projected sensitivity of PIXIE for $f_{\rm pbh} = 1$, but is one order of magnitude below the expected foreground $y$-parameter from the low-redshift intra-cluster medium \cite{Hill_15}. 

\textbf{UPDATE}

To conclude, we find that the global heating of the plasma due to accreting PBHs does not leave any observable imprint on CMB spectral distortions, neither for current instruments, nor for proposed ones. 

\subsubsection{Distortion from local Compton cooling}

There is another source of energy injection in the CMB, which occurs in the immediate vicinity of the PBH: when Compton cooling is efficient, the volumetric rate of energy transfer from the gas to the CMB is
\barr
\frac{d\dot{E}}{4 \pi r^2 dr} =  n_{\rm H} \frac{4 \overline{x}_e \sigma_{\rm T} \rho_{\rm cmb} }{m_e c (1 + \overline{x}_e)}(T - T_{\rm cmb}) \nonumber\\
\approx -\frac32 n_{\rm H} \rho^{2/3} v \frac{d}{dr}(T_{\rm cmb}/\rho^{2/3}),
\earr
where the second equality is obtained by setting $T \approx T_{\rm cmb}$ in the left-hand-side of Eq.~\eqref{eq:heat-compt}, which holds as long as Compton cooling is efficient. Therefore the rate of energy injection per PBH is 
\barr
\dot{E} &=& T_{\rm cmb} \int_{r_{\min}}^{\infty} dr 4 \pi r^2 v~ n_{\rm H} \frac{d (\ln \rho)}{dr}\nonumber\\
&=& \frac{\dot{M}}{m_p} T_{\rm cmb} \log(\rho(r_{\rm min})/\rho_{\infty}),
\earr
where $r_{\rm min} \sim \gamma^{-2/3} r_{\rm B}$ is the radius at which Compton cooling becomes inefficient, $\gamma$ being the dimensionless Compton cooling parameter defined in Eq.~\eqref{eq:gamma-def}. With $\rho(r_{\rm min}) \approx \rho_{\infty} (r_{\rm min}/r_{\rm B})^{-3/2}$, we arrive at
\beq
\dot{E} \sim \frac{\dot{M}}{m_p} T_{\rm cmb} \log(\gamma).
\eeq
We therefore get a characteristic distortion amplitude
\barr
\frac{\dot{\rho}_{\rm inj}}{H \rho_{\rm cmb}}  &\sim& f_{\rm pbh} \frac{\dot{M}}{H M} \frac{T_{\rm cmb} \rho_{\rm dm}}{\rho_{\rm cmb} m_p} \log(\gamma) \nonumber\\
&\sim& f_{\rm pbh} \frac{\dot{M}}{H M} \frac{n_{\rm H}}{n_{\rm cmb}},
\earr
where we have used $\rho_{\rm dm} \sim \rho_b$ and $n_{\rm cmb} \sim \rho_{\rm cmb}/T_{\rm cmb}$ is the number density of CMB photons. We see that this is proportional to the baryon-to-photon ratio $n_{\rm H}/n_{\rm cmb} \sim 10^{-10}$, and moreover multiplied by $H^{-1}\dot{M}/M$ which, as we discussed near Eq.~\eqref{eq:Mdot/HM}, is always less than unity for the mass range we consider. Therefore local Compton cooling by CMB photons does not lead to any observable spectral distortion.

%
%

\subsection{CMB temperature and polarization anisotropies}

\change{\subsubsection{Effect on CMB anisotropy power spectra}}

\begin{figure}
\includegraphics[width = \columnwidth]{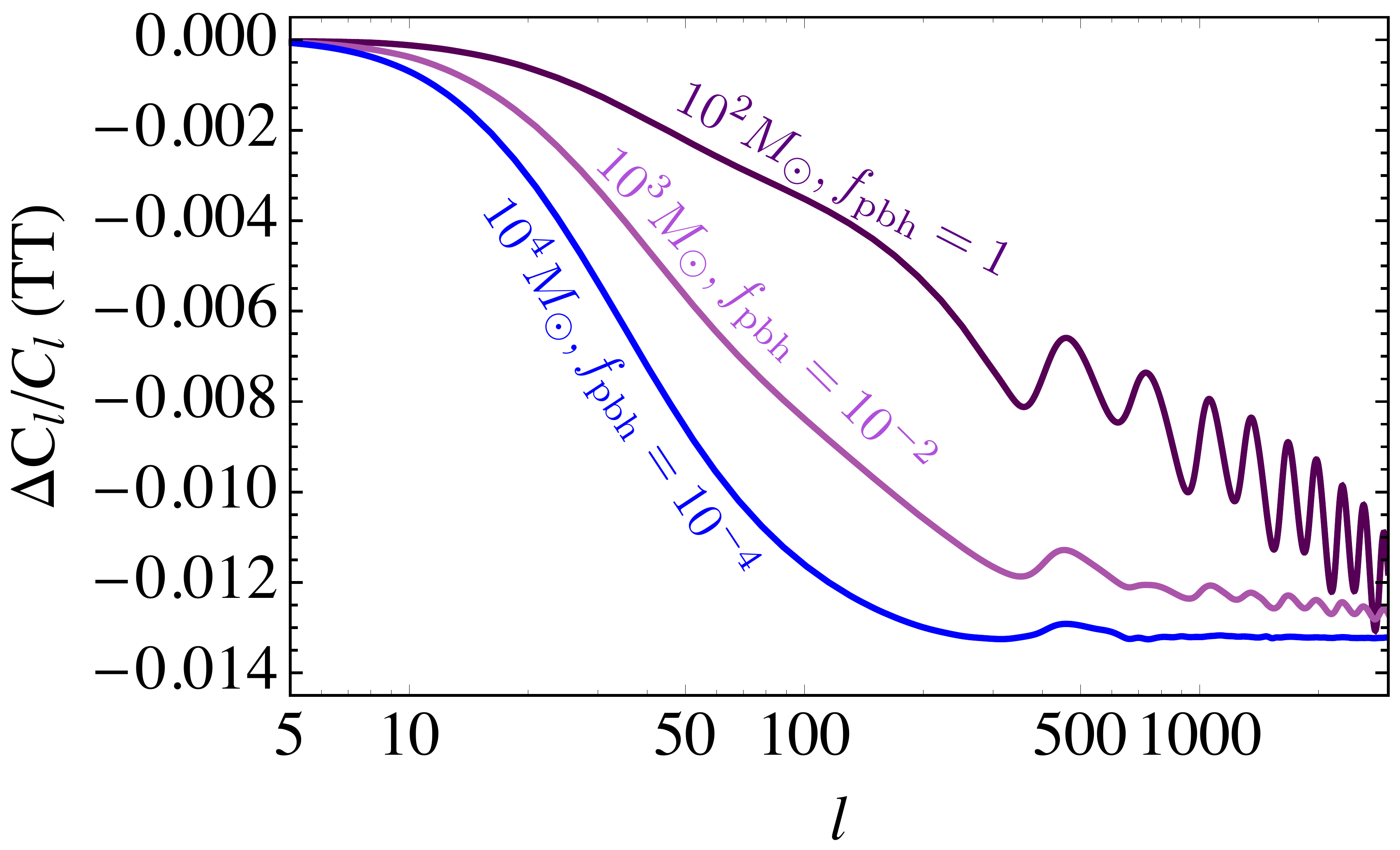}
\includegraphics[width = \columnwidth]{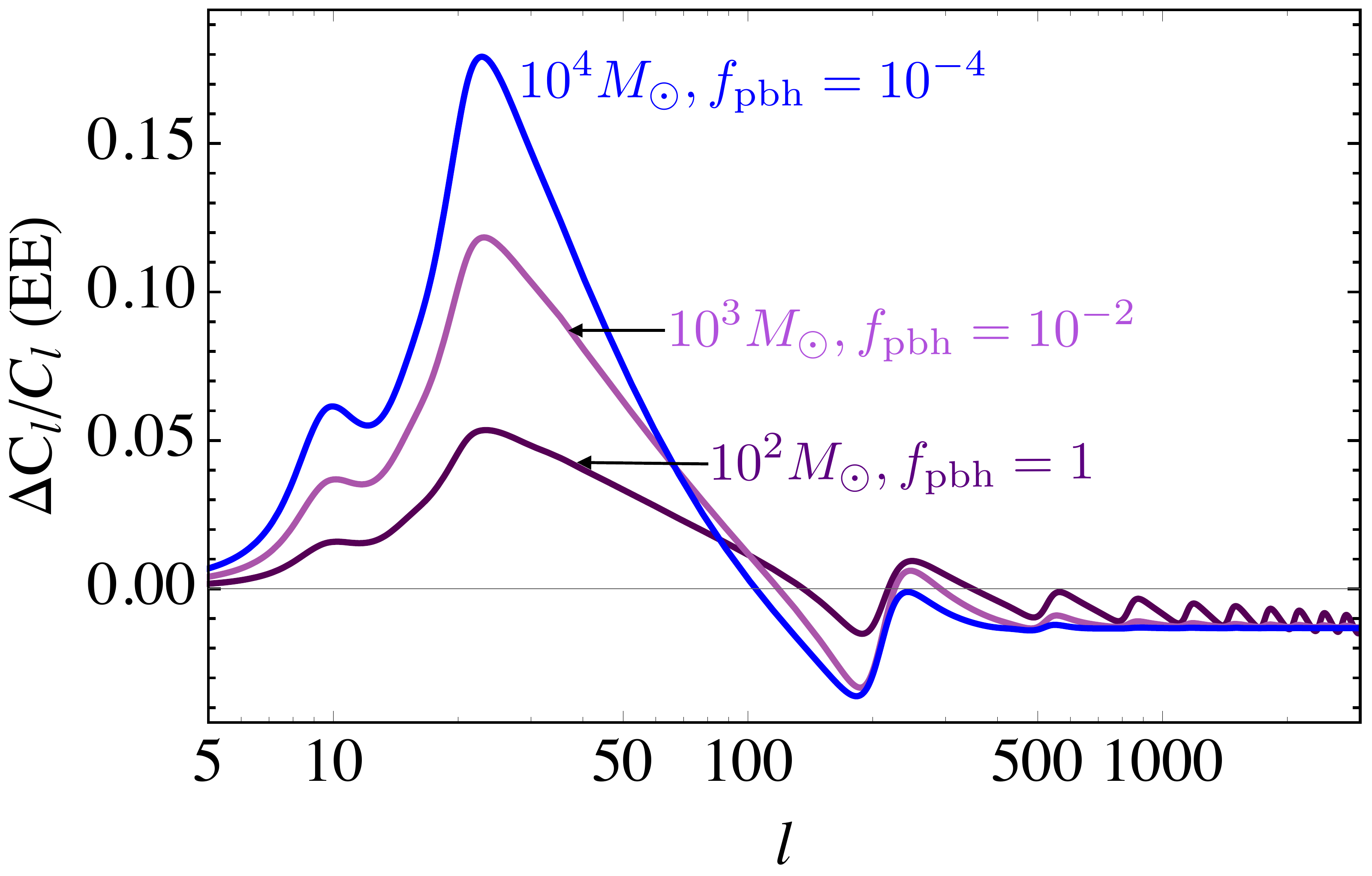}
\caption{Fractional change in the CMB temperature (\emph{upper panel}) and $E$-mode polarization (\emph{lower panel}) power spectra resulting from accreting PBHs. The parameters are $(M_{\rm pbh}/M_{\odot}, f_{\rm pbh}) = (10^2, 1), (10^3, 10^{-2}), (10^4, 10^{-4})$, in that order with increasing overall amplitude. We only show the \revision{collisional ionization} case here.} \label{fig:Cl}
\end{figure}

The change in the ionization history shown in Fig.~\ref{fig:xe} affects the visibility function for CMB anisotropies, and as a consequence the angular power spectra of temperature and polarization fluctuations. We have incorporated the modified \textsc{hyrec} into the Boltzmann code \textsc{class} \cite{Blas_11}. We show in Fig.~\ref{fig:Cl} the changes in CMB power spectra for the same parameters used in Fig.~\ref{fig:xe}. The effect is qualitatively similar to an increase in the reionization optical depth: fluctuations are damped on small angular scales due to scattering of photons out of the line of sight, and the polarization is enhanced on relatively large angular scales. The latter are smaller than the scales affected by reionization, as the effect of PBHs is at larger redshifts. For small PBH masses, the suppression on small scales is accompanied by oscillations, resulting from the change of the redshift of last scattering. Indeed, as can be seen in Fig.~\ref{fig:xe}, low-mass PBHs affect the recombination history near the last-scattering surface $z \sim 10^3$ more than high-mass PBHs, whose effect is mostly on the freeze-out free-electron fraction.

\change{\subsubsection{Analysis of Planck data}}

\change{To analyze the CMB anisotropy data from \emph{Planck}, one should in principle run a Monte Carlo Markov Chain (MCMC), accounting for foreground nuisance parameters (see e.g.~\cite{Planck_16}). However, this approach is too computationally taxing if we are to set an upper bound on the abundance of PBHs as a function of PBH mass, as it would require running a MCMC simulation for every mass considered. Instead, we performed a simplified yet accurate data analysis as follows.}

\change{We use the \texttt{Plik$\_$lite} best-fit $\hat{C}_{\ell}$ and covariance matrix $\bs{\Sigma}$ for the high-$\ell$ binned CMB-only $TT$, $TE$ and $EE$ power spectra provided by the \emph{Planck} collaboration\footnote{Available at http://pla.esac.esa.int/pla/} \cite{Planck_15_XI}. These spectra and their covariance matrix are obtained by marginalizing over foreground nuisance parameters. Since they are only provided for multipoles $\ell \geq 30$, we moreover assume a prior on the optical depth to reionization $\tau_{\rm reio} =  \tau_0 \pm \sigma_{\tau} \equiv 0.0596 \pm 0.0089$ as obtained by the latest \emph{Planck} data analysis \cite{Planck_16_XLVI}. This prior on $\tau_{\rm reio}$ accounts approximately \changee{for the large-scale temperature and polarization data (see Refs.~\cite{Munoz_16cip, Karwal_16} for an analysis similar in spirit)}. \changee{Given the relatively large effect of accreting PBHs on low-$\ell$ polarization (see Fig.~\ref{fig:Cl}), a full data analysis might change the constraints by order-unity factors; however this is below our theoretical uncertainty. For a given set of cosmological parameters $\vec{\theta} = (H_0, \Omega_b h^2, \Omega_c h^2, A_s, n_s, \tau_{\rm reio}, f_{\rm pbh})$ the $\chi^2$ is then}
\barr
\chi^2(\vec{\theta}) &=& \frac12 \left(C_{\ell}^X(\vec{\theta}) - \hat{C}^X_{\ell}\right) (\boldsymbol{\Sigma^{-1}})^{X X'}_{\ell \ell'}\left(C^{X'}_{\ell'}(\vec{\theta}) - \hat{C}^{X'}_{\ell'}\right)
 \nonumber\\
&+& \frac12 \frac{(\tau_{\rm reio} - \tau_0)^2}{\sigma_{\tau}^2},
\earr
where we sum over repeated indices, \changee{$X \in (TT, TE, EE)$, and the $C_{\ell}^{X}(\vec{\theta})$ are the theoretical power spectra} obtained with our modified \textsc{hyrec} and \textsc{class}. Taylor-expanding about the best-fit standard cosmological parameters $\vec{\theta}_0$ given in Ref.~\cite{Planck_16_XLVI} (with $f_{\rm pbh, 0} = 0$), we rewrite this as
\barr
\chi^2(\vec{\theta}) &\approx& \chi^2(\vec{\theta}_0) +  \Delta \theta_i  \frac{\partial C_{\ell}^X}{\partial \theta_i} \Big{|}_{\theta_0}  (\bs{\Sigma^{-1}})^{X X'}_{\ell \ell'}\left(C_{\ell'}^{X'}(\vec{\theta}_0) - \hat{C}^{X'}_{\ell'}\right)\nonumber\\
&+& \frac12 \Delta \theta_i F_{ij} \Delta \theta_j ,
\earr
where $\Delta \theta_i \equiv \theta_i -\theta_{0, i}$ and
\barr
F_{\ij} \approx \frac{\partial C^X_{\ell}}{\partial \theta_i} (\bs{\Sigma^{-1}})^{X X'}_{\ell \ell'}\frac{\partial C^{X'}_{\ell'}}{\partial \theta_j} +  \frac{\delta_{i, i_{\tau}} \delta_{j, i_\tau}}{\sigma_{\tau}^2}
\earr 
is the Fisher-information or curvature matrix \cite{Jungman:1995bz}, for which we have neglected the smaller term linear in $(C^X_{\ell}(\vec{\theta}_0) - \hat{C}^X_{\ell})$. Maximizing this quadratic approximation of the $\chi^2$ allows us to find the best-fit cosmological parameters $\hat{\vec{\theta}}$, with their covariance given by $(F^{-1})_{ij}$. We have checked that without PBHs this simple analysis recovers very accurately the best-fit standard 6 cosmological parameters obtained in Ref.~\cite{Planck_15_XI}, with biases of at most $0.17\,\sigma$. The variances we derive match those of Ref.~\cite{Planck_16} for $H_0, \Omega_b h^2, \Omega_c h^2$ and  $n_s$ and those of Ref.~\cite{Planck_16_XLVI} for $A_s$ and $\tau_{\rm reio}$, as expected since we are using the same high-$\ell$ covariance as in the former reference, and the prior on $\tau_{\rm reio}$ (strongly degenerate with $A_s$) from the latter.}

\change{We apply this analysis to derive the best-fit and 1-$\sigma$ error on $f_{\rm pbh}$, as a function of $M_{\rm pbh}$.} We explicitly checked that for the
limits we obtain, the change in the anisotropy power spectra is
indeed linear in $f_{\rm pbh}$ (ROM find an effect that goes as
$f_{\rm pbh}^{1/2}$ because they obtain a much larger effect on
the freeze-out free-electron abundance than we do). Though we consider a limit on the
normalization of a Dirac-function mass distribution, 
this analysis can be generalized to any extended
mass function \cite{Carr:1975qj}, by replacing $f_{\rm pbh}
L(M)/M \rightarrow \int
dM \frac{d f_{\rm pbh}}{d M}L(M)/M$, where $df_{\rm pbh}/dM$ is
the differential DM-PBH fraction.  

\begin{figure*}
\includegraphics[width = 1.3\columnwidth]{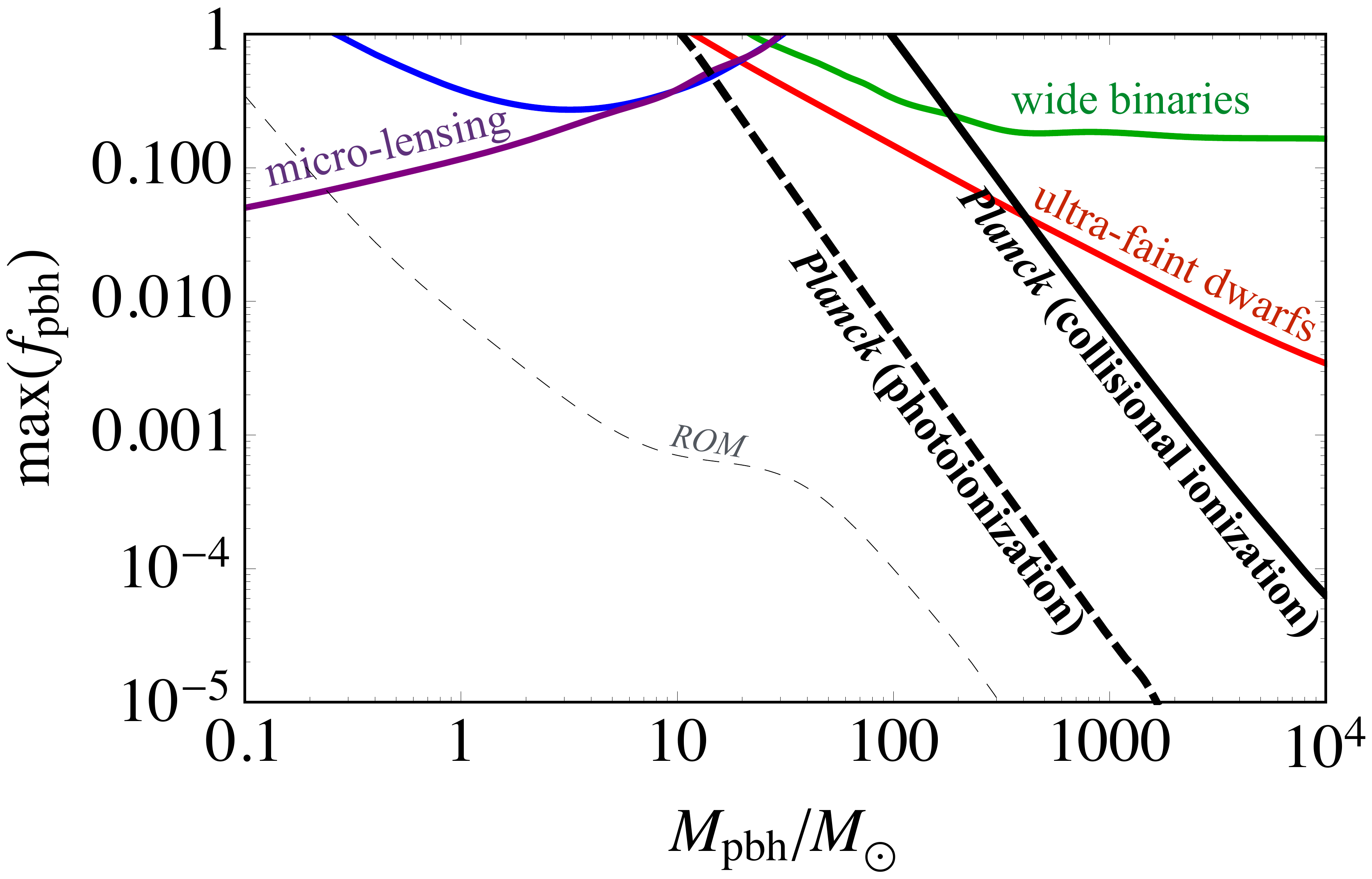}
\caption{Approximate CMB-anisotropy constraints on the fraction of dark matter made of PBHs derived in this work (\emph{thick black curves}). The \revision{``collisional ionization" case assumes that the radiation from the PBH does not ionize the local gas, which eventually gets collisionally ionized. The ``photoionization" case assumes that the local gas is ionized due to the PBH radiation, up to a radius larger than the collisional ionization region, yet smaller than the Bondi radius. The former case is the most conservative, as collisional ionization leads to a smaller temperature near the black hole horizon, hence a smaller luminosity, and weaker bounds. The correct result lies somewhere between these two limiting cases.} For comparison, we also show the CMB bound previously derived by ROM (\emph{thin dashed curve}), as well as various dynamical constraints: micro-lensing constraints from the EROS \cite{Eros_07} (\emph{purple curve}) and MACHO \cite{Macho_01} (\emph{blue curve}) collaborations \revision{(but see Ref.~\cite{Hawkins_15} for caveats)}, limits from Galactic wide binaries \cite{Monroy_14}, and ultra-faint dwarf galaxies \cite{Brandt_16} (in all cases we show the most conservative limits provided in the referenced papers).} \label{fig:constraints}
\end{figure*}

\change{For all PBH masses we consider, $M \leq 10^4 \,M_{\odot}$ (as the steady-state approximation breaks down beyond that mass), the best-fit $\hat{f}_{\rm pbh}$ is always less than a fraction of standard deviation\footnote{The astute reader may wonder why even given several probed PBH masses, some best-fit $\hat{f}_{\rm pbh}$ do not deviate by more than one standard deviation from 0; the reason is that the effect of PBHs of different masses on the CMB is very similar, hence the best-fit values are expected to be correlated. } $\sigma_{f_{\rm pbh}}$. We show $\sigma_{f_{\rm pbh}}$ in Fig.~\ref{fig:constraints}, as a simple proxy for the upper limit on this parameter\footnote{Strictly speaking, given the prior $f_{\rm pbh} \geq 0$, defining the 68\%-confidence interval is a bit more subtle; given the large uncertainties of the calculation, we shall not delve into such technical details here.}.} We see that
in the \revision{collisional ionization limit}, CMB anisotropy measurements by \emph{Planck} exclude PBHs with masses $M
\gtrsim 10^2\, M_{\odot}$ \change{as the dominant component of} the dark matter. \revision{In
the photoionization limit, this threshold is lowered to
$\sim 10\, M_{\odot}$. In either case}, our bound is significantly
weaker than that of ROM. The up to two orders of magnitude
difference in the constraint \revision{between the two limiting cases illustrates the level of uncertainty in the calculation. Nevertheless, we believe our most conservative
bound is robust and difficult to evade, at least at the order-of-magnitude level}.

\section{Discussion and conclusions} \label{sec:conclusion}

In this work we have revisited and revised existing CMB limits
to the abundance of primordial black holes. We showed that
CMB-anisotropy measurements by the \emph{Planck} satellite
\change{exclude PBHs as the dominant component of dark matter for masses $\gtrsim\, 10^2\, M_{\odot}$.}
The physical
mechanism involved is that PBHs would radiate a fraction of the
rest-mass energy they accrete, heating up and partially
reionizing the Universe. Such an increase in the free-electron
abundance would change the CMB temperature and polarization
power spectra. \emph{Planck} measurements do not allow for large
deviations from the standard recombination history
\cite{Planck_16}, which leads to tight bounds for large and
luminous PBHs.

The constraints we derive are significantly weaker than the
previous result of Ricotti et al.~(ROM) \cite{Ricotti_08}, so it
is instructive to briefly summarize the differences in our
respective calculations. First and foremost, we compute the
radiative efficiency $\epsilon \equiv L/\dot{M} c^2$ from first
principles, generalizing Shapiro's classic calculation for
spherical accretion around a black hole \cite{Shapiro_73}. We
account for Compton drag and cooling as well as ionization
cooling once the background gas is neutral. At fixed accretion
rate, the efficiency we derive is at least a factor of ten and
up to three orders of magnitude lower than what is assumed in
ROM for spherically-accreting PBHs. The second largest
difference is in the accretion rate itself. ROM compute the
accretion rate for an isothermal equation of state, assuming
that Compton cooling by CMB photons is always very efficient. In
fact, for sufficiently low redshift and low PBH masses Compton
cooling is negligle and the gas is adiabatically \change{heated}. In this case
the higher gas temperature, and hence pressure, imply an
accretion rate that is lower by a factor of $\sim 10$.
Since the PBH luminosity is
quadratic in the accretion rate, this translates to a factor of
$\sim 100$ reduction in the effect of PBHs on CMB observables. A
third difference is the relative velocity between PBHs and
baryons, which ROM significantly underestimates around $z\sim
10^3$, leading to an over-estimate of the accretion
rate.

\change{There are considerable
theoretical uncertainties in the calculation of the accretion
rate and luminosity of PBHs, as we have illustrated by considering two limiting cases for the radiative feedback on the local ionization fraction, leading to largely different results. Let us recall the most critical uncertainties here}. First, we have only
considered spherical accretion. Extrapolating the measured
primordial power spectrum to the very small scale corresponding
to the Bondi radius, ROM estimated the angular momentum of the
accreted gas; they argued that the accretion is indeed spherical
for PBHs less massive than $\sim 10^3- 10^4\, M_{\odot}$. However,
there is no direct measurements of the ultra-small-scale power
spectrum, and all bets are open for a Universe containing
PBHs. If small-scale fluctuations are larger (for instance due
to non-linear clustering of PBHs), an accretion disk could form,
with a significantly enhanced luminosity with respect to
spherical accretion. On the other hand, non-spherical accretion
could conceivably also lead to complex three-dimensional flows
near the black hole giving rise to a turbulent pressure that
\change{lowers the accretion rate and radiative output}.
Secondly, we have accounted for the motion
of PBHs with an approximate and very uncertain rescaling of the
accretion rate. Given that dark-matter-baryon relative
velocities are typically supersonic, we expect shocks and a much
more complex accretion flow in general. Thirdly, we have assumed
a steady-state flow, but have not established whether such a
flow is stable, even for a static black hole. Last but not
least, if PBHs only make a fraction of the dark matter, an
assumption must be made about the rest of it, the simplest one
being that it is made of weakly interacting massive particles
(WIMPs). If so, these WIMPs ought to be accreted by PBHs, whose
mass may grow significantly after matter-radiation equality
\cite{Mack_07}, and as a consequence increase the accretion rate
of baryons \cite{Ricotti_07, Ricotti_08}. For the sake of simplicity, and given the added
uncertainty associated with the accretion of collisionless
particles, we have not accounted for this possibility in this
work. \changee{Given these major qualitative uncertainties, we have made several simplifications leading to additional factors of a few inaccuracies: for instance, our calculation is purely Newtonian, and our analytic treatments of the ionization region and of energy deposition into the plasma are only approximate. We have also only explicitly analyzed \emph{Planck}'s temperature and polarization data for multipoles $\ell \geq 30$, approximating the effect of large-scale measurements by a simple prior on the optical depth to reionization. In a nutshell, the reader should keep in mind that this is a complex problem and that many simplifying assumptions underly our results, which we expect to be accurate at the order-of-magnitude level only.}

\revision{To conclude, we find that, up to the theoretical uncertainties aforementioned, CMB anisotropies conservatively rule out PBHs more massive than $\sim 10^2\, M_{\odot}$ as the dominant form of dark matter. This bound could be tighter by up to one order of magnitude if the local gas is predominantly photoionized rather than collisionally ionized. Given the recent renewed interest in the $\sim 10-100 \, M_{\odot}$ window, it would be very interesting to generalize our accretion model to self-consistently account for ionization feedback, a task beyond the scope of this article, and to be pursued in future work. In the mean time,} there are a number of other interesting astrophysical probes in that mass range. These include
future measurements of
the stochastic gravitational-wave background \cite{Pen:2015qta,Mandic:2016lcn,Cholis:2016xvo,Clesse:2016ajp,Wang:2016ana} and of the
mass spectrum \cite{Kovetz:2016kpi}, redshift distribution \cite{Nakamura:2016hna}, and orbital eccentricies \cite{Cholis:2016kqi} for future
binary-black-hole mergers; lensing of
fast radio bursts by PBHs \cite{Munoz:2016tmg}; pulsar timing \cite{Schutz:2016khr,Inomata:2016rbd}; \change{radio/x-ray sources \cite{Gaggero:2016dpq}} or the cosmic
infrared background \cite{Kashlinsky:2016sdv}; the dynamics of
compact stellar systems \cite{Brandt_16}; strong-lensing systems
\cite{Mediavilla:2009um}; and perhaps
clustering of GW events
\cite{Clesse:2016vqa,Namikawa:2016edr,Raccanelli:2016cud, Raccanelli_16}. The
conclusions of our work suggest that it will be important to
pursue vigorously these alternative avenues.

%

\section*{Acknowledgments}
We are grateful to Graeme Addison, Juli\'{a}n Mu\~{n}oz and Vivian Poulin for helpful conversations about the \emph{Planck} likelihood, Fisher analysis, and energy deposition, respectively. \revision{We also thank Massimo Ricotti for insightful feedback on this work}. We thank Simeon Bird, William Dawson, Alvise Raccanelli and Pasquale Serpico for useful comments on this manuscript. We also acknowledge conversations with
Jens Chluba, Ilias Cholis, S\'{e}bastien Clesse, Juan
Garc\'{i}a-Bellido, Ely Kovetz, Julian Krolik, Katie Mack, and
Hong-Ming Zhu. This work was supported by the Simons
Foundation, NSF grant PHY-1214000, and NASA ATP grant
NNX15AB18G. 

\appendix

\section{Consistency checks} \label{app: consistency}

\subsection{Isolated PBH assumption} \label{app:isolated}

Our calculation assumes gas accreting on an isolated BH. This approximation is valid as long as the Bondi radius is significantly smaller than the characteristic proper separation $\overline{r}$ between PBHs. Numerically, we get
\barr
r_{\rm B} &=& \frac{G M}{v_B^2} \approx 6 \times 10^{14} \textrm{cm} \frac{M}{M_{\odot}} \frac{10^3}{1+z}\\
\overline{r} &=& \left(\frac{3 M}{4 \pi f_{\rm pbh} \overline{\rho}_{\rm dm}}\right)^{1/3} \nonumber\\
&\approx& 6 \times 10^{17} \textrm{cm}  \left(\frac{M}{f_{\rm pbh} M_{\odot}}\right)^{1/3}  \frac{10^3}{1+z},
\earr
where we estimated the Bondi radius for a PBH at rest and assuming $T_{\rm gas} = T_{\rm cmb}$ and $x_e \ll 1$, valid for $200 \lesssim z \lesssim 1100$. We therefore find that the isolated PBH approximation holds for
\beq
M \lesssim 3 \times 10^4 f_{\rm pbh}^{-1/2} M_{\odot}.
\eeq
Given that our conservative bound is $f_{\rm pbh} \lesssim (100\, M_{\odot}/M)^2$, PBHs can indeed be considered as isolated for all masses considered. 

Note, however, that this estimate only holds for quasi-uniformly distributed PBHs. In practice PBHs may cluster significantly if they make up a significant fraction of the dark matter, due to Poisson fluctuations in their initial clustering \cite{Afshordi_03}. We do not attempt to account for this effect here.

%
%
%
%

\subsection{Free-Free cooling} \label{sec:free-free cooling}

Free-free cooling is efficient when the associated timescale $t_{\rm ff} \sim n_e T/j_{\rm ff}$ is much shorter than the local accretion timescale $t_{\rm acc} \sim r/|v|$. The ratio of the two timescales is
\beq
\frac{t_{\rm acc}}{t_{\rm ff}}  \sim \frac{r/|v|}{n_e T/j_{\rm ff}}\sim \frac{\alpha c \sigma_{\rm T} n_e r}{|v|}\mathcal{J} \sim \frac{\alpha c \sigma_{\rm T} \dot{M}}{4 \pi m_p r v^2} \mathcal{J},
\eeq
where we have used $n_e = \rho/m_p = \dot{M}/(4 \pi m_p r^2 |v|)$. In the innermost region, the gas is in near free-fall, so that $r v^2 \sim GM$. Recalling that the Eddington luminosity is given by Eq.~\eqref{eq:Ledd}, we may rewrite this as
\beq
\frac{t_{\rm acc}}{t_{\rm ff}}  \sim \dot{m} \alpha \mathcal{J}.
\eeq
Therefore, as long as $\dot{m} \lesssim$ a few, we may safely neglect free-free cooling. This is indeed the case for the mass range $M \lesssim 10^4 M_{\odot}$ that we consider (see Fig.~\ref{fig:lambda}).

\subsection{Free-bound radiation} \label{sec:fb}

At low frequencies, near the ionization threshold of hydrogen, radiative recombinations also contribute to the radiation of the plasma \cite{Shapiro_73}. We consider only recombinations to the ground state of hydrogen, for which the cross-section is well known and has the approximate dependence near threshold
\beq
\sigma_{\rm pi}(\nu) \approx \sigma_0 \left(\frac{\nu_{\rm I}}{\nu}\right)^3,
\eeq
with $\sigma_0 \approx 6 \times 10^{-18}$ cm$^2$ and $\nu_{\rm I} \equiv E_{\rm I}/h$ is the threshold photoionization frequency. Assuming Saha equilibrium, detailed balance considerations allow us to compute the corresponding free-bound emissivity (see e.g.~Ref.~\cite{Osterbrock_06}): 
\beq
j_{\nu}^{\rm fb} =  n_e^2  (3 \pi m_e T)^{-3/2} \frac{8 \pi h^4 \nu^3}{c^2} \rme^{- h (\nu - \nu_{\rm I})/T}\sigma_{\rm pi}(\nu).
\eeq
Therefore the free-bound emissivity near threshold is nearly independent of frequency\footnote{This result differs from Shapiro's assumed free-bound spectrum.}. The ratio of free-bound to free-free emissivities is 
\beq
\frac{j^{\rm fb}}{j^{\rm ff}} \sim \left(\frac{h \nu_{\rm I}}{\sqrt{m_e c^2 T}}\right)^3 \frac{\sigma_0}{\alpha \sigma_{\rm T}} \ll 1.
\eeq
Even though $\sigma_0 \gg \alpha \sigma_{\rm T}$, this ratio is largely suppressed due to the first factor. 

\subsection{Optical thickness} \label{sec:opt_thin}

In our estimate of the luminosity we have assumed that the plasma is optically thin. Here we show that the plasma is indeed optically thin to both Compton scattering and free-free absorption.

The Compton optical depth is dominated by the densest regions near the horizon. Since the Compton cross section is lower than Thomson for relativistic photons, the Compton optical depth is less than
\beq
\tau_{\rm Com} \lesssim r_{\rm S} n_e \sigma_{\rm T} \sim \dot{m},
\eeq
where we used Eqs.~\eqref{eq:ne} and \eqref{eq:Ledd}. Therefore, as long as $\dot{m} \lesssim 1$ the plasma is optically thin to Compton scattering \cite{Ricotti_08}.

The free-free absorption coefficient $\alpha_{\nu}^{\rm ff}$ (with dimensions of inverse length) is \cite{Rybicki_86}
\beq
\alpha_{\nu}^{\rm ff} = \frac{j_{\nu}^{\rm ff}}{B_{\nu}(T)},
\eeq
where $j_{\nu}^{\rm ff} \approx \frac{j^{\rm ff}}{4 \pi} h/T$ is the emissivity and $B_{\nu}(T)$ is the Planck function. Since $j_{\nu}^{\rm ff} \propto n_e^2$ the total optical depth is dominated by the region near the horizon. The optical depth is then
\beq
\tau^{\rm ff} \sim r_{\rm S} \alpha_{\nu}^{\rm ff} \sim r_{\rm S} \frac{j c^2 }{h \nu_{\max}^4},
\eeq
where we approximated $j_{\nu} \sim j/\nu_{\max}$ and $B_{\nu} \sim h \nu_{\max}^3/c^2$, where $\nu_{\max} \sim T_{\rm S}/h$. Using Eq.~\eqref{eq:j_ff}, we get
\beq
\tau_{\rm ff} \sim \alpha ~\tau_{\rm Th} \left(\frac{h c}{T_{\rm S}} n_e^{1/3}\right)^3. \label{eq:tau_ff}
\eeq 
The last term is the degeneracy factor: if it is greater than unity one ought to account for electron degeneracy pressure. It is easy to check that this term is always much smaller than unity for all cases considered.

\section{Energy deposition rate: comparison with the existing literature} \label{app:comparison}

Several papers attempt an analytic estimate of the energy deposition rate as we do in Section \ref{sec:deposition}, as opposed to a fully numerical treatment as in Ref.~\cite{Slatyer_09}. Here we compare and contrast our results to the existing literature. Reference~\cite{Cirelli_09} gives the following integral expression for the photon density per energy interval [their equation (2.12), rewritten in our notation]: 
\barr
\mathcal{N}_E(t) &=& \int^t dt' \frac1{E'} \frac{d \dot{\rho}_{\rm inj}}{dE'} \left(\frac{a'}{a}\right)^3 \nonumber\\
&&\times\exp\left[- \int_{t'}^t dt'' n_A'' c \sigma_{\rm
tot}(E'')\right],
\label{eq:twotwelve}
\earr
where $E' \equiv E a/a'$, $E'' \equiv E a/a''$, $n_A$ is the number density of absorbers and $\sigma_{\rm tot}(E)$ is total cross section for all the interactions suffered by the DM-sourced photon and that result in the production of free electrons. This integral equation is equivalent to the following partial differential equation:
\beq
a^{-3}\frac{d}{dt}(a^3 \mathcal{N}_E) = \frac1{E} \frac{d \dot{\rho}_{\rm inj}}{dE} - n_A c \sigma_{\rm tot}(E) \mathcal{N}_E. \label{eq:Cirelli}
\eeq
This equation differs from our Eq.~\eqref{eq:continuity} in two ways. First, in the absence of photon sources or sinks, Eq.~\eqref{eq:Cirelli} does not recover the correct scaling $\mathcal{N}_E \propto a^{-2}$. Second, the second term on the right-hand side implies that photons are destroyed in the ionization process. While this is the case for direct photoionization events $\gamma + H \rightarrow p + e$, it is not the case for ionizations following Compton scattering events $\gamma + H \rightarrow p + e + \gamma'$, for which part of the energy of the incoming photon is used for ionizing the atom, but the photon is not destroyed.

From Eq.~(\ref{eq:twotwelve}), Ref.~\cite{Poulin_15} deduces the energy deposition rate (correcting a mistake in Refs.~\cite{Natarajan_10, Giesen_12}). Assuming $\sigma_{\rm tot}(E) \approx \sigma_{\rm T}$ (valid for $E \lesssim m_e c^2$), the resulting energy deposition rate is (Eq.~(4.21) of Ref.~\cite{Giesen_12}, corrected in Appendix B of Ref.~\cite{Poulin_15})
\barr
&&\dot{\rho}_{\rm dep} = \int^t dt' \rme^{- \kappa(t, t')} \left(\frac{a'}{a}\right)^8 \overline{n}_{\rm H}'c \sigma_{\rm T} \dot{\rho}_{\rm inj}' \ \ (\textrm{Poulin et al.}) \nonumber,\\
&&\textrm{with} \ \ \kappa(t, t') \equiv \int_{t'}^t d t'' \overline{n}_{\rm H}'' c \sigma_{\rm T}.
\earr
Rewriting this as differential equation would lead to
\beq
a^{-8} \frac{d (a^8 \dot{\rho}_{\rm dep})}{dt} = \overline{n}_{\rm H} c \sigma_{\rm T} \left[\dot{\rho}_{\rm inj} - \dot{\rho}_{\rm dep} \right].
\eeq
This differs from our Eq.~\eqref{eq:rho_dep} in two ways. First, the incorrect scaling $\dot{\rho}_{\rm dep} \propto a^{-8}$ instead of $a^{-7}$ once energy deposition becomes inefficient is a direct consequence of the incorrect scaling in Eq.~(2.12) of Ref.~\cite{Cirelli_09}. Secondly, our right-hand-side is smaller by an (approximate) factor $0.1$. This translates the fact that, even for $E \lesssim m_e c^2$, only a small fraction of the energy of Compton-scattered photons is lost to ionizations, as opposed to the totality of it as implicitly assumed in Eq.~\eqref{eq:Cirelli}.

\bibliography{pbh_cmb}

\end{document}